\documentclass[twocolumn,apj,floatfix]{aastex631}
\pdfoutput=1
\usepackage{amsmath,amssymb,gensymb,color,graphicx}
\usepackage{array}
\usepackage{natbib,hyperref}
\usepackage{multirow}
\usepackage{placeins}
\usepackage{textcomp}
\begin{document}


\title{The Atacama Cosmology Telescope: Quantifying Atmospheric Emission above Cerro Toco}

\author{Thomas~W.~Morris}
\email{thomas.w.morris@yale.edu}
\affiliation{Department of Physics, Yale University, New Haven, CT 06511, USA}
\affiliation{National Synchrotron Light Source II, Brookhaven National Laboratory, Upton, NY 11973, USA}

\author{Elia~Battistelli}
\affiliation{Sapienza University of Rome, Physics Department, Piazzale Aldo Moro 5, 00185 Rome, Italy}

\author{Ricardo~Bustos}
\affiliation{Departamento de Ingeniería Eléctrica, Universidad Católica de la Santísima Concepción, Alonso de Ribera 2850, Concepción, Chile}

\author{Steve~K.~Choi}
\affiliation{Department of Physics and Astronomy, University of California, Riverside, CA 92521, USA}

\author{Adriaan J. Duivenvoorden}
\affiliation{Center for Computational Astrophysics, Flatiron Institute, New York, NY 10010, USA}
\affiliation{Joseph Henry Laboratories of Physics, Jadwin Hall, Princeton University, Princeton, NJ 08544, USA}

\author{Jo~Dunkley}
\affiliation{Joseph Henry Laboratories of Physics, Jadwin Hall, Princeton University, Princeton, NJ 08544, USA}
\affiliation{Department of Astrophysical Sciences, Peyton Hall, Princeton University, Princeton, NJ 08544, USA}

\author{Rolando D\"unner}
\affiliation{Instituto de Astrofísica and Centro de Astro-Ingeniería, Facultad de Fìsica, Pontificia Universidad
Católica de Chile, Av. Vicuña Mackenna 4860, 7820436 Macul, Santiago, Chile}

\author{Mark Halpern}
\affiliation{Department of Physics and Astronomy, University of British Columbia, Vancouver, British Columbia V6T 1Z1, Canada}

\author{Yilun~Guan}
\affiliation{Dunlap Institute for Astronomy \& Astrophysics, University of Toronto, 50 St. George St., Toronto ON M5S 3H4, Canada}

\author{Joshiwa~van~Marrewijk}
\affiliation{European Southern Observatory, Karl-Schwarzschild-Straße 2, 85748 Garching bei München, Germany}

\author{Tony~Mroczkowski}
\affiliation{European Southern Observatory, Karl-Schwarzschild-Straße 2, 85748 Garching bei München, Germany}

\author{Sigurd~Naess}
\affiliation{Institute for Theoretical Astrophysics, University of Oslo, Sem Sælands vei 13, 0371 Oslo, Norway}

\author{Michael D. Niemack}
\affiliation{Department of Physics, Cornell University, Ithaca, NY 14853, USA}
\affiliation{Department of Astronomy, Cornell University, Ithaca, NY 14853, USA}

\author{Lyman~A.~Page}
\affiliation{Joseph Henry Laboratories of Physics, Jadwin Hall, Princeton University, Princeton, NJ 08544, USA}

\author{Bruce~Partridge}
\affiliation{Department of Physics and Astronomy, Haverford College,
Haverford, PA 19041, USA}

\author{Roberto Puddu}
\affiliation{Instituto de Astrofísica and Centro de Astro-Ingeniería, Facultad de Fìsica, Pontificia Universidad
Católica de Chile, Av. Vicuña Mackenna 4860, 7820436 Macul, Santiago, Chile}

\author {Maria Salatino}
\affiliation{Stanford University, Stanford, CA 94305, USA}
\affiliation{Kavli Institute for Particle Astrophysics and Cosmology, Stanford, CA 94305, USA}

\author{Crist\'obal~Sif\'on}
\affiliation{Instituto de F\'isica, Pontificia Universidad Cat\'olica de Valpara\'iso, Casilla 4059, Valpara\'iso, Chile}

\author{Yuhan~Wang}
\affiliation{Department of Physics, Cornell University, Ithaca, NY 14853, USA}

\author{Edward~J.~Wollack}
\affiliation{NASA/Goddard Space Flight Center, Greenbelt, MD, USA 20771}

\begin{abstract}

At frequencies below 1\,Hz, fluctuations in atmospheric emission in the Chajnantor region in northern Chile are the primary source of interference for bolometric millimeter-wave observations.
This paper focuses on characterizing these fluctuations using measurements from the Atacama Cosmology Telescope (ACT) and the Atacama Pathfinder Experiment (APEX) water vapor radiometer. We show that the total precipitable water vapor (PWV) is not in general an accurate estimator of the level of fluctuations in millimeter-wave atmospheric emission. We also show that the microwave frequency spectrum of atmospheric fluctuations is in good agreement with predictions by the \texttt{am} code for frequency bands above 90~GHz. We introduce a new method for separating atmospheric and systematic fluctuations, allowing us to fit a robust atmospheric flatfield, as well as to study in the atmosphere in greater detail than previous works. We present a direct measurement of the temporal outer scale of turbulence of $\tau_0\approx50$\,s corresponding to a spatial scale of $L_0\approx500$\,m. Lastly, we show the variance of fluctuations in ACT's mm-wave bands correlate with the variance of fluctuations in PWV measured by APEX, even though the observatories are 6\,km apart and observe different lines of sight. 

\end{abstract}

\section*{}
\clearpage
\newpage

\section{Introduction}

For CMB observations at $90$ GHz and above, a large source of noise at frequencies below $1\,$Hz is from spatial fluctuations in atmospheric water vapor as it blows through the field of view of the telescope. If these fluctuations can be better modeled and removed from the time stream, it may be possible to measure larger angular scales than are currently achieved from the ground.
In a step towards this goal, \citet{morris/etal:2022} showed we could determine the bulk atmospheric wind velocity above the Atacama Cosmology Telescope (ACT) using the time-dependent correlation function of the detectors within one of ACT's detector arrays. 
This paper extends these results in a number of ways. We show that the characteristic spatial structure of turbulence extends well beyond the field of view of the ACT receiver (three arrays over $2^\circ$) and that it scales between frequencies following the predictions of the \texttt{am} model \citep{am:2019}. By separating the effects of the telescope scanning from intrinsic atmospheric fluctuations over large angular and temporal scales, we measure the temporal atmospheric structure function and compute the outer scale of turbulence, $L_0$: 
%
in the ideal case, atmospheric fluctuations as measured by CMB experiments over timescales less than an hour are fully described by the fluctuation amplitude, outer scale, and wind velocity. In addition to wind and water vapor density, air temperature also fluctuates with a power law spectrum \citep{obukhov:1949}, but has a different form near the inner scale of turbulence\footnote{A much smaller scale than what ACT measures.} \citep{champagne/etal:1977, williams/paulson:1977, hill/clifford:78}. We see possible evidence of these temperature fluctuations at 30 and 40 GHz.

The theoretical foundation for quantifying atmospheric turbulence is given in \citet{tatarski} and references therein. Building on those results, \citet{church:1995} gave a framework for predicting the fluctuation levels in CMB experiments. Later, \citet{lay/halverson:2000} developed and applied a moving frozen sheet (2D) model to compare turbulent emission between the South Pole and Chilean sites. \citet{errard/etal:2015} extended \citet{church:1995} to directly compare data from POLARBEAR, which is next to ACT, to a parameterized 3D model of turbulence. The \texttt{TOAST}\footnote{\url{https://github.com/hpc4cmb/toast}} software follows 
the \citet{errard/etal:2015} model to predict the sensitivity of CMB experiments, while the \texttt{maria}\footnote{\url{https://github.com/thomaswmorris/maria}} software \citep{van2024maria} uses the two-dimensional approximation introduced in \citet{morris/etal:2022}.

The approach here and in \citet{morris/etal:2022} complements \citet{errard/etal:2015} in that we seek specific signatures in the data that correspond to specific physical aspects of the atmosphere such as the wind speed and outer scale of turbulence.
Although we do not consider polarized mm-wave emission, first measured by \citet{Troitsky_Osharin:2000}, we note recent advances linked to specific mechanisms: \citet{petroff/etal:2020} measured circular polarization from the Zeeman splitting of O$_2$'s magnetic dipole transitions, and \citet{takakura/etal:2019}, \citet{li2023class} and \citet{coerver2024spt} measured linear polarization from scattering off ice crystals in clouds.

After a brief overview of mm-wave emission and turbulence in the Chajnantor region in 
Section~\ref{sec:chajnantor-atmosphere},
we outline a statistical formulation of atmospheric turbulence in Section~\ref{sec:stat_desc}. Then in Section~\ref{sec:apex}, we give an overview of the APEX weather station followed, in 
Section~\ref{sec:instrument}, by
a brief description of relevant features of the ACT receiver. Section~\ref{sec:processing} describes the pre-processing of the raw data needed to reliably quantify the atmospheric signal. Based on this, Section~\ref{sec:results} gives results from the ACT data, presenting measurements of the power spectrum and structure function of atmospheric turbulence along with its characteristic outer scale. In addition, we compare ACT's fluctuation levels to those measured by APEX. We conclude in Section~\ref{sec:discussion}.

\section{Atmospheric conditions in the Chajnantor region} 
\label{sec:chajnantor-atmosphere}

In the millimeter-wave regime, the atmospheric emission spectrum is dominated by oxygen (with a pair of spectral lines at 60 and 119 GHz) and water vapor (with spectral lines at 22, 183, 321, 325, and 380 GHz, shown in Figure~\ref{fig:atmosphere_spectrum_with_passbands}). The atmosphere's water content is typically quantified with the precipitable water vapor, or the column depth of water directly above the observer in units of length.
We refer to the zenith water column depth as $w_z(t)$, and more generally to the line-of-sight water vapor (which depends on zenith angle) as $w(t)$. While the oxygen content is fairly constant, the amount of water vapor fluctuates on both short timescales (less than an hour) due to turbulent mixing and long timescales (hours or longer) due to weather and climatological effects. Our focus is on the former regime.

\begin{figure}[htb]
\includegraphics[width=0.475\textwidth]{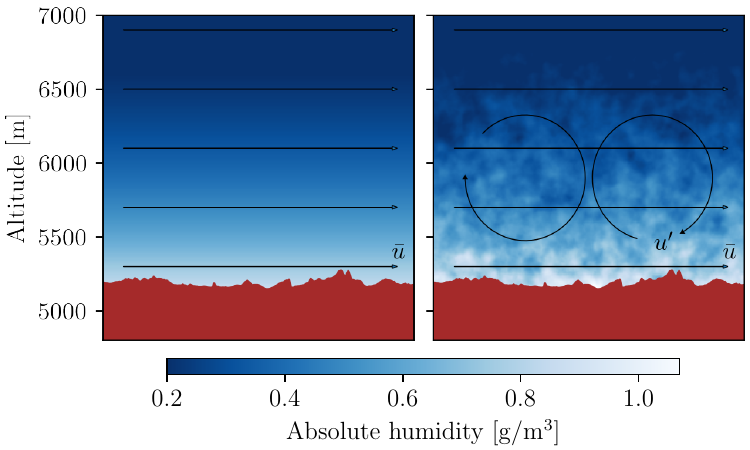}
\caption{\textit{Left:} a vertical gradient in absolute humidity with constant horizontal wind velocity $\bar{u}$. \textit{Right:} a gradient with turbulent vortices $u^\prime$ arising from the wind shear
that transport parcels of air between different layers. For both plots, the mean water vapor as a function of height is the same. The amplitude of fluctuations depends on the strength of the gradients of the wind and water vapor.}
\label{fig:pict_of_turb}
\end{figure}

Water vapor fluctuates on short timescales due to turbulent mixing. Consider a turbulent vortex spanning a gradient in humidity, for example between a wetter layer of air closer to the ground and a dryer layer farther up in the atmosphere. The velocity field will transport air from the wetter layer to the dryer layer and vice versa, leading to turbulent variations in water vapor content. This is depicted in Figure~\ref{fig:pict_of_turb} and described in \cite{tatarski} as turbulent diffusion.

The degree of turbulence is proportional to the (primarily vertical) gradient of the quantity of interest. The quantities with appreciable gradients are wind velocity, water
vapor content compared to the mean fraction and in some cases
temperature variation compared to the mean vertical profile, such as
when there is a thermal inversion layer.\footnote{In absence of a temperature inversion,
the typical gradient is $-6.5^\circ$/km.} Oxygen and nitrogen have
comparatively smaller gradients compared to nominal composition. Because of its strong emission lines, water vapor fluctuations are particularly important for mm-wave measurements, but we can think of them as a marker for the turbulence that affects all components of the atmosphere. In the language of \cite{tatarski}, water vapor is a ``passive additive.'' \citet{obukhov:1949} shows that temperature fluctuations can also be treated in the framework of a passive additive.

\section{Quantifying turbulent fluctuations}
\label{sec:stat_desc}

In the Kolmogorov picture of turbulence, kinetic energy is injected into the flow velocity by wind shear at some outer scale $L_0$ in the form of turbulent vortices. These gradually cascade into smaller and smaller vortices until they inhabit some length scales $l_0$ small enough for viscous forces to be appreciable, at which point the kinetic energy is dissipated as heat. For a velocity field in three dimensions, the organization of kinetic energy at different length scales follows a scale-invariant spectrum\footnote{This form corresponds to the three-dimensional power spectrum. As only the magnitude of $\kappa$ is relevant, it can be expressed as a one-dimensional spectrum in which case $\Phi(\kappa) \propto \kappa^{-5/3}$. Both forms lead to the ``2/3'' dependence in the structure function.}

\begin{equation}
\label{eqn:kolmogorov}
    \Phi(\kappa) \propto \kappa^{-11/3}, \quad L_0^{-1} < \kappa < l_0^{-1},
\end{equation}
where $\kappa$ is the wavenumber. If we posit a negligibly small inner scale and a flat spectrum for spatial frequencies smaller than $L_0^{-1}$, we can model the spectrum as
\begin{equation}
\label{eqn:modified-kolmogorov}
    \Phi(\kappa) \propto (\kappa^2 + L_0^{-2})^{-11/6}.
\end{equation}
This corresponds to a covariance function 
\begin{equation}
\label{eqn:turbulent-covariance}
    C(r) = \sigma^2 M_{1/3}\bigl(\sqrt{2/5}~ r/L_0\bigr),
\end{equation}
where $r$ is a spatial separation, $\sigma^2$ is the variance of the fluctuations, and $M_{\nu}(r)$, 
the normalized Mat\'ern covariance function (Appendix~\ref{appendix:matern}),
contains the spatial dependence.
Equation~\ref{eqn:turbulent-covariance}
is the same as Equation 1.33 in \cite{tatarski} with $r_0=L_0/\sqrt{2/5}$, but is more compact and more broadly used 
(e.g., \citet{matern:1986}, \citet{genton:2002}).
It is often more convenient to work with the structure function, $D(r) = \big \langle [ y(x + r) - y(x) ]^2 \big \rangle = 2 \big ( C(0) - C(r) \big )$, where $y(x)$ is a stationary random function with zero mean,\footnote{Note that $\langle y(x)^2\rangle =C(0)$ and $\langle y(x+r)y(x)\rangle =C(r)$. }  which applied to Equation~\ref{eqn:turbulent-covariance} yields the Kolmogorov 2/3 power law. For small separations, $r \ll L_0$, we have (see Appendix~\ref{appendix:matern})
\begin{equation}
\label{eqn:two-thirds-law}
    D(r) \approx C_0^2 r^\frac{2}{3}=5.028\sigma^2(r/L_0)^\frac{2}{3} .
\end{equation}
Tatarski shows that the distribution of a passive additive turbulently diffused by such a velocity field follows the same form, meaning that Equation~\ref{eqn:turbulent-covariance} applies also to the turbulent mixing of water vapor.
%

\subsection{Turbulent distributions as seen by an observer on the ground}

To a detector observing the sky, fluctuations in emission will appear as the integral along the line-of-sight through the three-dimensional distribution described above.
The total power in each detector is determined by integrating over the contribution from each element in the beam. Under the assumption that the atmosphere is optically thin and the emission is proportional to the water vapor content of the atmosphere, we can, for notational simplicity, represent the total factor of the response of detector $i$ to the atmosphere as a single gain $g_i$ such that its fluctuation in power $\delta P_i$ is given by
\begin{equation}
    \delta P_i(t) = g_i \delta w_i(t),
\end{equation}
where $\delta w_i(t)$ is the time-varying total water vapor in the line-of-sight of detector $i$.  Typical values for $g$ for a $45^\circ$ elevation are given in Table~\ref{tab:atmpwr}. 

In Appendix~\ref{appendix:matern}, we present the correlation function measured by a telescope observing through the atmosphere including the effects of the beams. 
To quantify an outer scale, we adopt the frozen transport model \citep[e.g.,][]{lay, lay/halverson:2000}. 
This leads to a correspondence, $r = u\tau$, between spatial separation $r$ and time delay $\tau$, where $u$ is a horizontal wind velocity.
In support of this approximation, \citet{morris/etal:2022} showed that the integral of the 3D wind velocity (from reanalysis data products like ERA5 \citep{ERA5}) times a 3D turbulence model through the atmosphere agrees well with a single wind speed multiplied by a 2D model of turbulence. In essence, the spatial pattern of turbulence is transported at velocity $u$ at some effective height.
For timescales greater than a second, a good approximation 
(Appendix~\ref{appendix:two-dimensional}) is
\begin{equation}
\label{eqn:time-covariance}
    C(\tau) \approx \sigma^2M_{5/6}(uL_0^{-1}\tau),
\end{equation}
with the corresponding structure function
\begin{equation}
\label{eqn:time-sf}
    D(\tau) = 2 \big [ C(0) - C(\tau) \big ] = 2\sigma^2 [1 - M_{5/6}(uL_0^{-1}\tau)]
\end{equation}
that for $\tau \ll uL_0^{-1}$ becomes
\begin{equation}
\label{eqn:time-sf-power-law}
    D(\tau) \propto \tau^{5/3}.
\end{equation}
The corresponding power spectrum is the Fourier transform of Equation~\ref{eqn:time-covariance}
\begin{equation}
\label{eqn:time_ps_power_law}
    \Phi(f) \propto (f^2 + f_0^2)^{-8/6},
\end{equation}
that for $f \gg f_0$ has the approximation 
\begin{equation}
\label{eqn:time_ps_power_law_approx}
    \Phi(f) \propto f^{-8/3}.
\end{equation}

This one-dimensional approximation sufficiently describes atmospheric fluctuations measured by ACT, and is also used in e.g., \cite{consortini1972choice}, \cite{conan2000analytical} and \cite{tokovinin2002differential}. In contrast to the work here, \citet{errard/etal:2015} use a Gaussian correlation function for which $L_0$ is roughly comparable, as shown in Figure~\ref{fig:covariance-triptych}. 

\section{The APEX weather station}
\label{sec:apex}

\begin{figure}[htb]
\includegraphics[width=0.475\textwidth]{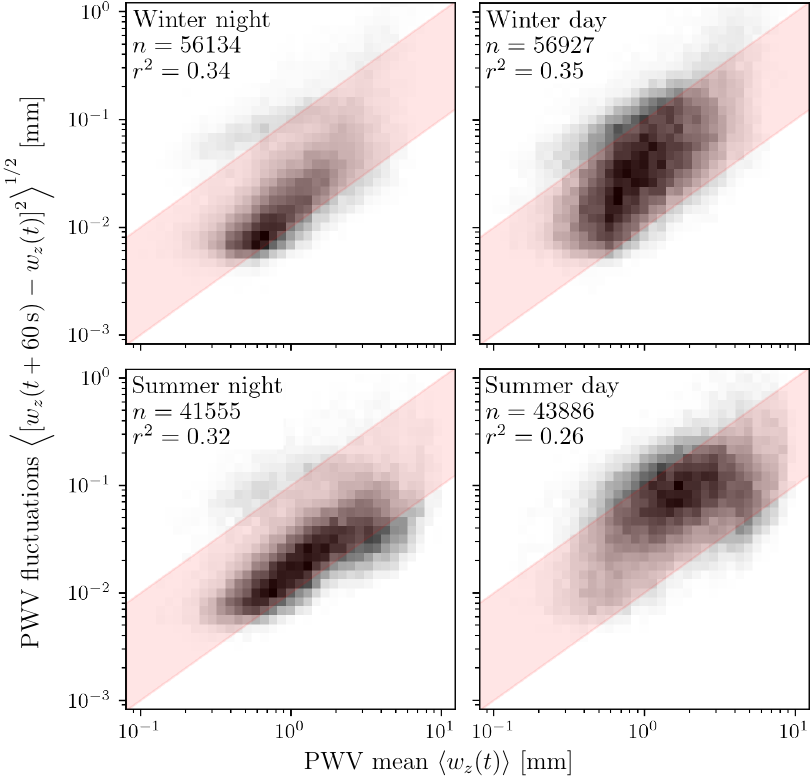}
\caption{Half-hourly mean $w_z$ versus fluctuations in $w_z$, as reported by APEX. We define the austral summer as the months from November to April, and the day as the hours between 11:00 and 23:00 UTC. Total water vapor is not a good predictor of fluctuation level, with a correlation coefficient in log-log space of $r^2=0.36$ for all data. The shaded red regions denote relative fluctuation levels of between 1\% and 10\% of the total, containing 76\% of the data.}
\label{fig:apex_pwv}
\end{figure}

\begin{figure}[htb]
\includegraphics[width=0.475\textwidth]{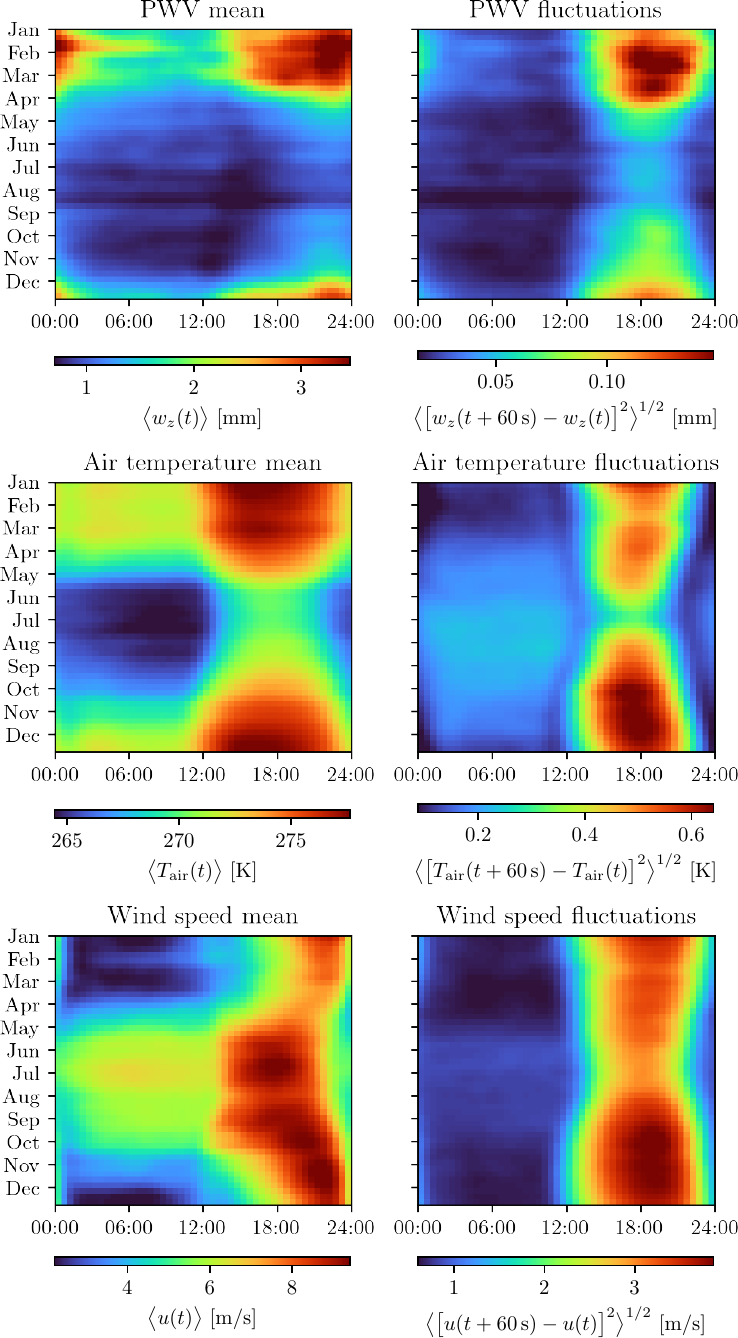}
\caption{The diurnal and seasonal trends in the mean and fluctuation level of several atmospheric components. The times on the $x$-axis are in UTC; local time is GMT-4.}
\label{fig:apex_weather_stats}
\end{figure}

The APEX telescope (5064\,m above mean sea level, 5107\,m GPS altitude, \citet{APEX_obs, APEX_rad1:Kerber_2015}) operates and makes freely available\footnote{\url{https://archive.eso.org/wdb/wdb/asm/meteo_apex/form}} minute-by-minute weather data, including $w_z$ deduced from their 183~GHz water vapor radiometer. The radiometer follows the design of \citet{rose/etal:2005} and was built by RPG Radiometer Physics GmbH \citep{rose/czekala:2006,sarazin/kerber/debreuck:2013}. It can output an average every second, but APEX reports just one of these values per minute due to the need to combine the raw radiometer output with meteorological data to determine 
$w_z$ \citep{carlos}.\footnote{The radiometer uses International Atomic Time (TAI) which currently leads UTC by 37 seconds to the accuracy needed in this work.} 
The resolution of the measurements is 0.01 mm. The APEX WVR is 82\,m lower and 6\,km away from ACT, and varies in both azimuth and elevation. 
This different line of sight means that fluctuations in PWV on the order of a few minutes are not phase coherent between ACT and APEX. However, the amplitude of these fluctuations are well correlated on timescales of several hours, as we show below.

The scale height of water vapor at APEX is typically 1309\,m \citep[][Table 5]{cortes/etal:2016}; see also \citet{APEX_rad2}.
However, \citet{radford/etal:2008} note that it varies considerably and can be as low as 750\,m at night. \citet{bustos/etal:2014} directly compare measurements by APEX and a similar radiometer on the top of Cerro Chajnantor (elevation 5612~m) and also show evidence in support of an inversion layer between 5064~m and 5612~m. Based on these 
values, $w_z$ for ACT should be 0.9 to 0.94 times that of APEX.

Figure~\ref{fig:apex_pwv} shows a summary of all APEX radiometer measurements taken between January 2006 and May 2024. The $y$-axis gives the median value and the fluctuation level obtained by removing a running mean from the time-ordered data with a window of 30 minutes and computing the standard deviation of the readings in 30-minute bins.
Typically, the RMS is a few percent of $w_z$, but there is considerable variation as we elaborate on below. The plot shows that on average, a summer night has smaller fluctuations than a winter day; in general, winter is drier than summer and days have both more moisture and relative fluctuation than nights.
Figure~\ref{fig:apex_weather_stats} shows the means and fluctuations in $w_z$, temperature, and wind speed from APEX. Again, this shows that nights are favorable over days. It also shows that relative to their totals, PWV fluctuations are much stronger than temperature fluctuations; typically by a factor of ten or more. \citet{otarola/etal:2019} present a complementary summary. 


While APEX's measurements of $w_z$ represent the entire atmospheric column and largely agree with ACT's, its measurements of wind and temperature do not, likely because they sample only the surface.
Specifically, \cite{morris/etal:2022} shows that APEX wind measurements do not fit with ACT's determination of the wind, which is based on the movement of the full column of water. Similarly, APEX's temperature measurements do not agree with ERA's. Nevertheless, we use the relative magnitude of the fluctuations to set our expectations. 

\section{The atmosphere as seen by ACT}
\label{sec:instrument}
\subsection{The ACT instrument}
\label{sec:act_inst}

ACT was\footnote{ACT was decommissioned in October 2022 to make way for the Simons Observatory \citep{SO:2019}.} a 6\,m mm-wave telescope located in the Chajnantor region of northern Chile at an altitude of 5145\,m above mean sea level (GPS altitude, 5188\,m). There were three generations of receivers on the ACT telescope \citep{fowler/etal:2007,swetz/etal:2011}. In this paper we analyze data from the last, Advanced ACTPol (AdvACT) \citep{henderson/etal:2016,ho/2017,choi/etal:2018,li/etal:2018,li/etal:2021}, which used NIST-built polarization-sensitive detectors operating at 100\,mK \citep{grace/etal:2014,datta/etal:2014,ho/etal:2016}. 

For a complete description of the receiver, we refer the reader to \citep{thornton/etal:2016}; here we give the salient features relevant to this work. The roughly 1\,m diameter receiver holds three independent, hexagonal, approximately $400$-feed polarization-sensitive detector arrays (PAs), each at the base of a customized ``optics tube" that couples the array to the telescope. Looking at the face of the receiver, as shown in \citep{thornton/etal:2016}, these optics tubes are arranged in a triangle with a center typically near $45^\circ$ elevation. When projected onto the sky, the combined beam pattern of all three tubes has a triangular pattern with a base of about $2^\circ$. At each vertex of the triangle is the beam pattern for one of the hexagonal arrays. The edge-to-edge diameter of the hexagonal pattern is about $0.75^\circ$ as shown in Figure 2 of \citet{murphy/etal:2024}. The entire telescope scans the arrays with an azimuthal speed of $1.5^\circ/$sec.


AdvACT comprised the PA4, PA5, PA6, \& PA7 dichroic arrays that took the place of ACTPol's PA1, PA2, and PA3 \citep{thornton/etal:2016}. The PA4 array measures near 150 and 220 GHz, which we designate as f150 and f220 bands, while the PA5 and PA6 arrays each measure near 90 and 150 GHz, designated as f090 and f150. The PA6 array is highest on the sky. In the last couple of years of ACT's operation PA6 was swapped out for the lower frequency PA7, operating in the f030 and f040 bands, that is described in Section VII.C below. 

\subsection{Predicted response to fluctuations in PWV}

The measured power in a frequency band is given by
\begin{equation}
P = \eta k_B\int T_b(\nu)\tau(\nu )d\nu,
\end{equation}
where the passband, $\tau(\nu )$, is the response of the receiver to a Rayleigh-Jeans (RJ) source and normalized to unity, $\eta$ is the overall efficiency of the telescope and receiver, and $T_b(\nu )$ is the brightness temperature of the atmosphere as a function of frequency as shown in 
Figure~\ref{fig:atmosphere_spectrum_with_passbands}. To model $T_b(\nu )$ we use
\texttt{am} (version 13.0, see \cite{am:2019}). The total power in each band, and its derivative with respect to the $w_z$, is given in Appendix~\ref{appendix:results} as a function of $w_z$ for $\eta=1$.

%
%
%

\begin{figure}[htb]
\includegraphics[width=.5\textwidth]{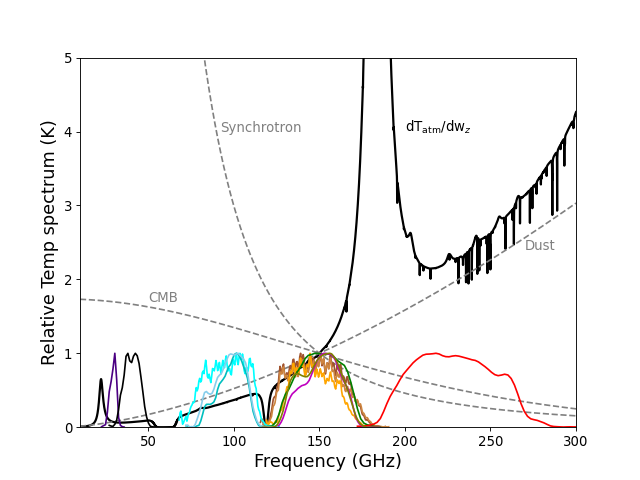}
\caption{The derivative of the atmospheric emission temperature with respect to $w_z$ for $w_z=1~$mm normalized at 150 GHz (solid black), computed with {\tt am} using median atmospheric parameters for Chajnantor. Derivatives at different $w_z$ are similar. The prominent feature is the water line 183 GHz. For comparison, we show the brightness temperature for synchrotron, CMB, and dust emission all normalized at 150 GHz. For reference, a fluctuation in $w_z$ of 0.1 mm in PA5 f150 corresponds to a change in brightness temperature of 0.64~K. The normalized passbands for PA1–7 are shown for reference.}
\label{fig:datm_dpwv}
\end{figure}

In the time stream, the primary quantity of interest near 90 and 150 GHz is the fluctuation in emission with respect to $w_z$, $\delta T_b(\nu)/\delta w_z$. This is shown in 
Figure~\ref{fig:datm_dpwv}, normalized at 150 GHz,  along with the spectra of other sources. The atmosphere has a steep spectrum and is potentially useful for constraining out-of-band leakage. 

It is more straightforward to work in units of power as opposed to brightness temperature
because the first step in ACT's calibration procedure is to convert the detector output to picowatts (pW). 
In response to fluctuations $\delta w_z$, assuming we receive a single mode of radiation, we measure power fluctuation
\begin{multline}
\label{eq:power-fluctuations}
    \delta P = \frac{\partial\,P_\mathrm{atm}}{\partial\,w_z}\delta w_z = \eta \left[ k_B\int \frac{ \partial\, T_b(\nu)}{\partial\,w_z}\tau(\nu )d\nu\right] \delta w_z \\= g_w\delta w_z ,
\end{multline}
where the quantity in brackets is given in Table~\ref{tab:atmpwr}
as $\partial P/\partial w_z$ for a zenith angle of 45$^\circ$. 
%
%
%
Analogously, the fluctuations in power resulting from fluctuations in atmospheric temperature are given by

\begin{equation}
\label{eq:temperature-fluctuations}
    \delta P = \eta \left[ k_B\int \frac{ \partial\, T_b(\nu)}{\partial\,T_\mathrm{air}}\tau(\nu )d\nu\right] \delta T_\mathrm{air}.
\end{equation}

\begin{figure}[htb]
\includegraphics[width=.5\textwidth]{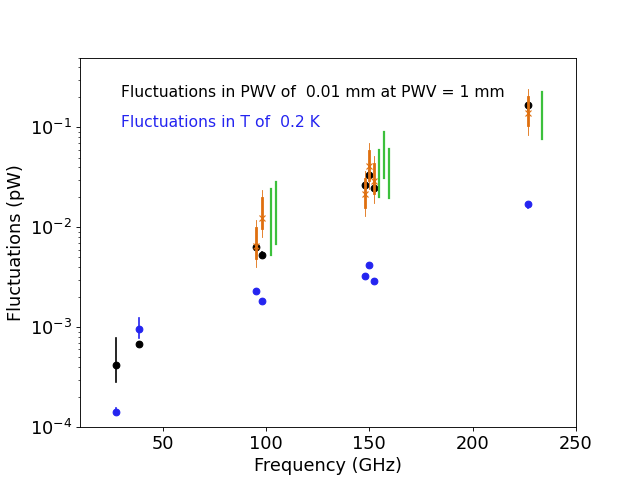}
\caption{The fluctuations in picowatts at the input of the receiver vs. frequency for two different models, one in which the source is water vapor density (Equation~\ref{eq:power-fluctuations}) with 
$\delta_{w_z}=0.01\,$mm,
typical for a winter night (e.g., Figure~\ref{fig:apex_weather_stats}), 
and the other in which the source is fluctuations in atmospheric temperature of 0.2~K (Equation~\ref{eq:temperature-fluctuations}). Under these conditions,
the dominant fluctuations for $f>90\,$GHz are from water vapor.
The error bars at 30 and 40 GHz show the range from shifting the passband $\pm 2$~GHz. The orange and green (offset by +7\,GHz) lines  are discussed in Section~\ref{sec:results}. Because of an absence of correlation with APEX as discussed in Section~\ref{sec:low-frequency}, we do not plot any data at 30 and 40 GHz.}
\label{fig:fluctuation_spectrum}
\end{figure}

The frequency spectrum of both kinds of fluctuations is shown in Figure~\ref{fig:fluctuation_spectrum}. Our data clearly favor fluctuations in water vapor density as opposed to temperature at $\geq 90$ GHz as evidenced below.

\section{Processing of ACT data}
\label{sec:processing}

\subsection{Time streams and pre-processing}

We consider ACT data taken during the 2017, 2018, and 2019 observing seasons (May 2017 to January 2020). We also consider observations at 30 and 40 GHz from the 2020 and 2021 seasons in Section~\ref{sec:low-frequency}.
We use the same raw data, cuts, and calibration that go into the mapping pipeline \citep{aiola/etal:2020, naess/etal:2020}.
ACT data typically consists of hour-long periods of observations between detector re-biasing, which are then divided into five approximately 11-minute long chunks of time-ordered data (each referred to as a ``TOD''). 
For each TOD, we down sample the data from 400\,Hz to 100\,Hz using a triangular filter (see e.g. \citep{dunner/etal:2013}). In analyzing the structure function, we remove a linear trend over multiple concatenated TODs to remove an offset and long-term trends in the data. We emphasize that the cosmological analysis corresponds to data at higher frequencies ($>0.5$ Hz) and smaller spatial scales ($\ell \gtrsim 200$ or $\lesssim 1^\circ$) than those used here.

\begin{figure*}[tp!]
\includegraphics[width=0.95\textwidth]{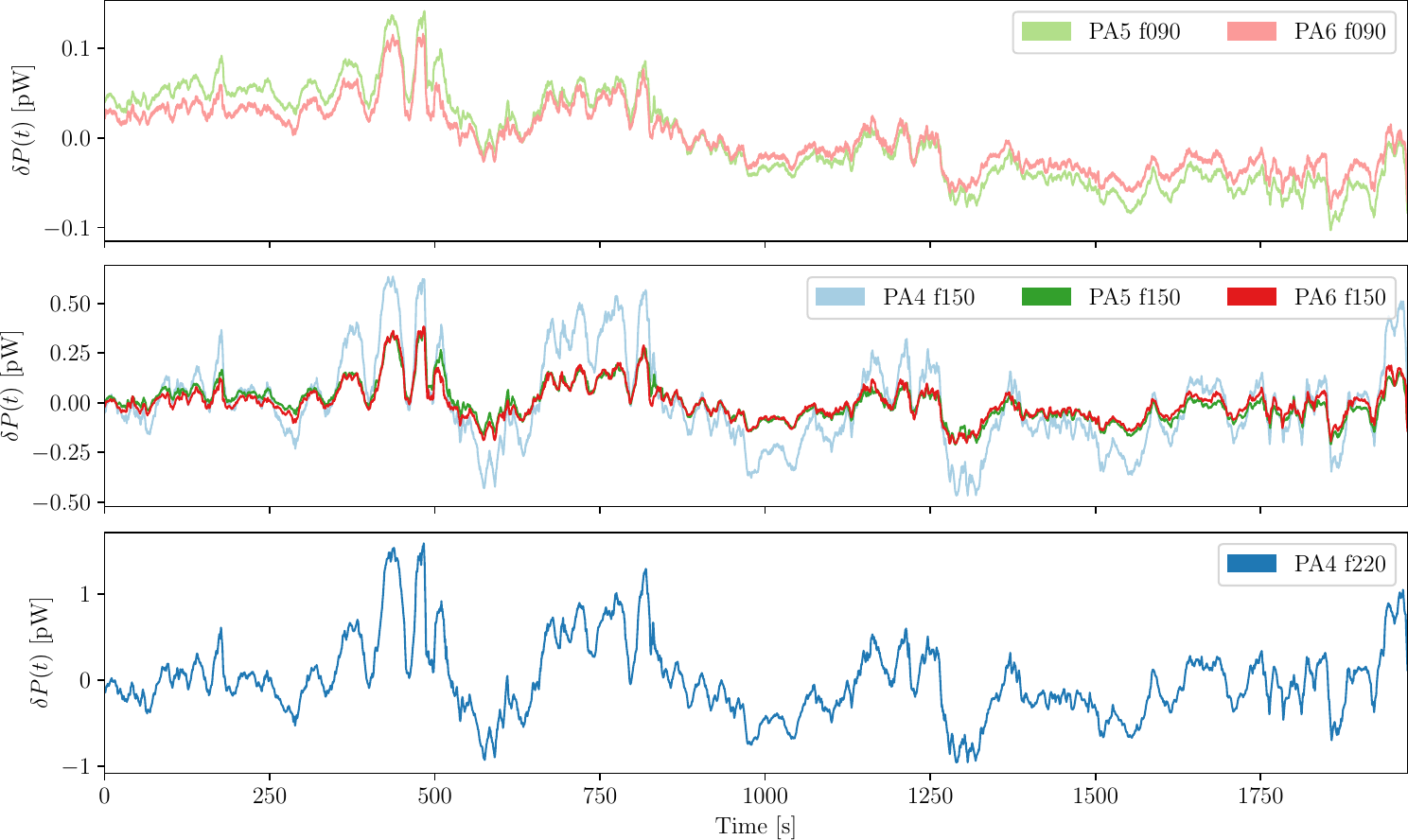}
\caption{Roughly 40 minutes of ACT data, with one representative detector timestream (with the mean subtracted) taken from each band for an observation at an elevation of $40^\circ$ starting at 19:24:50 UTC on August 15, 2019 when ${\rm PWV_{ACT}}=0.9\,$mm. Note the strong correlation between arrays. The different amplitudes at 150 GHz are due to different passbands and efficiencies.}
\label{fig:sample_tod}
\end{figure*}

\subsection{Mode separation and TOD selection}
\label{sec:tod-decomposition}

Figure~\ref{fig:sample_tod} shows a typical receiver output for 40 minutes. The common mode signal in all three optics tubes and their bands is from the atmosphere and readily apparent. 
However, there are often optics tube-common modes that do not correspond to emission from the atmosphere or sky; these modes are periodic and depend on both scan position \textit{and} direction, and thus clearly not on the sky. We attribute these to systematic pickup for each optics tube (hereafter ``pickup''). When atmospheric fluctuations are weak, scan-synchronous pickup can become comparable to or even dominate atmospheric fluctuations, confounding analysis of the atmosphere. Binning and subtracting out a template of left- and right-going scans is common method of removing this pickup for CMB analysis, but it is less useful for atmospheric analysis because it may remove some of the atmosphere. We therefore want a method that separates atmospheric fluctuations and systematic pickup.

To minimize assumptions, we posit only that the fluctuations, $\delta P$, for detector $i$ in optics tube $j$ can be modeled as
\begin{equation}
    \delta P_{i,j} = g_{w, i} \delta w + \sum_j g_{p, i} \delta p_j + \epsilon_i(t),
\end{equation}
where $g_{w, i}$ and $g_{p, i}$ are the responses of detector $i$ to the fluctuations in water vapor ($\delta w$) and pickup in optics tube $j$ ($\delta p_j$), respectively, and $\epsilon_i(t)$ is some residual.
The pickup modes are common to an optics tube, with each detector in that tube having a different gain to it,\footnote{Based on a phenomenological examination of the data, we find that a linear response of each detector to the underlying systematic mode in each optics tube is sufficient.}
while the atmosphere is common to all three optics tubes.\footnote{Fluctuations smaller than ACT's field of view are negligibly small compared to the common mode.}
 With one common atmospheric mode and three optics tube pickup modes, we represent the ($n_\text{detector} \times n_\text{time}$) matrix of all detector fluctuations $\delta \mathbf{P}$ for each TOD, $t$, as
\begin{equation}
\label{eqn:tod-decomposition}
    \delta \mathbf{P}_t = \mathbf{G}\,\delta \mathbf{M}_t.
\end{equation}
Here, $\mathbf{G}$ is a ($n_\text{detector} \times 4$) matrix of gains and $\mathbf{M}$ is a ($4 \times n_\text{time}$) matrix consisting of the stacked modes $\{\delta w, \delta p_1, \delta p_2, \delta p_3 \}$. If we assume that all detectors have zero response to pickup on adjacent optics tubes (see Figure~\ref{fig:gain-matrix}), then we have 6 measurements (each band) per 4 unknowns (each mode).

We fit Equation~\ref{eqn:tod-decomposition} over all data, allowing $\mathbf{G}$ to vary slowly as a cubic spline with a knot spacing of 90 days. The fitting procedure is described in Appendix~\ref{appendix:tod-decomposition}. Among other considerations, for each TOD it condenses the matrix $\delta \mathbf{M}$ into a single $(4 \times 4)$ matrix $\mathbf{X}$ that preserves the correlations between different modes but is much smaller. We take all TODs for which there were at least 16 detectors from each band, representing around 90\% of the total.
For each detector in each TOD, we remove a cubic spline with a knot spacing of ten minutes to separate the matrix of power fluctuation $\delta \mathbf{P}$ from the total power $\mathbf{P}$ and to limit the contribution of low frequency modes from the fitting of the gain matrix. This spline is used only to identify atmospheric modes and not for their subsequent analysis. Per TOD (averaged over all detectors), the relative residual $||\delta \mathbf{P}_t - \mathbf{G}\,\delta \mathbf{M}_t|| / ||\delta \mathbf{P}_t||$ has a median of $3\%$, suggesting that for most TODs this model successfully describes the fluctuations in all detectors.\footnote{When using only one common mode and no pickup modes, the median relative residuals are on the order of $100\%$.}

After fitting, we obtain a unique gain matrix $\mathbf{G}$ for each season and an ``embedding'' matrix $\mathbf{X}$ (with a vector representing each mode) for each TOD. This decomposition is useful for diagnosing the relative amount of atmospheric and non-atmospheric fluctuations in the timestreams for each detector. The fitted $\mathbf{G}$ is also useful on its own for constructing a flat field that is resistant to systematic pickup. A robust flat-field is essential for avoiding calibration-dependent power loss \citep{naess/louis:2023}. Note that there is one overall scaling degeneracy between $\mathbf{G}$ and $\mathbf{X}$ that cannot be uniquely 
determined by
the data. Later, we show that we can calibrate the fitted $\mathbf{G}$ by comparing the scale of the embedding of the common mode in $\mathbf{X}$ to independent measurements of the strength of water vapor fluctuations, i.e. the APEX radiometer.

Given the separation of each TOD into common and non-common modes, we are able to estimate the relative strength of atmosphere and non-atmospheric fluctuations. For each detector-TOD in our later atmospheric analysis, we define the per-TOD atmospheric ``purity''
\begin{equation}
    a^2 \equiv \sigma^2_\text{c} / \sigma^2_\text{nc},
\end{equation}
where $\sigma^2_\text{c}$ and $\sigma^2_\text{nc}$ are the variances of the common and non-common modes in each detector's signal. Distributions of $a^2$ for each band are shown in Figure~\ref{fig:a-squared_histogram}.
We set the threshold for a detector-TOD being ``atmosphere-dominated'' as $a^2 \geq 10$, that is, when the variance of the atmospheric fluctuations is at least 10 times that of systematic fluctuations.
We restrict the data to detector-TODs with this designation to minimize systematic contamination in quantifying the characteristics of atmospheric turbulence. 

\subsection{Structure functions}
\label{sec:coangular-structure-functions}

After the decomposition performed in Section~\ref{sec:tod-decomposition}, we have an estimate of $a^2$ for each detector-TOD. In probing the outer scale of fluctuations, it is helpful to have as long a string of data as possible; thus in computing structure functions and estimating outer scales, we concatenate three TODs for a total of 33 minutes of observation, and remove a linear trend.\footnote{We omit the first TOD after each calibration, as sometimes the detectors are not yet in thermal equilibrium as ACT begins to scan.} We keep only 3-TOD chunks that have an average $a^2$ of $\geq 10$, such that the entire chunk can be said to be dominated by atmospheric fluctuations.

ACT employs a wide back-and-forth constant-elevation raster scan, meaning that the structure function describing its data has both a temporal and azimuthal dependence. To sample the temporal structure function, we take 256 logarithmically-spaced samples of $\tau$ between $0.1$ and $1000$ seconds and for each $\tau$ find all pairs of times $t_i, t_j$ such that $t_i - t_j = \tau$ and $\phi(t_i) = \phi(t_j)$, where $\phi(t)$ is the azimuth of the boresight as a function of time. The structure function at $\tau$ is then the average of square differences $D(\tau) \langle [\delta P(t_i) - \delta P(t_j)]^2  \rangle_\textrm{pairs}$. To probe the structure function at levels below the white noise of individual detectors, we take as $\delta P$ the average power of all detectors for each band.



\subsection{Estimating the coefficient of turbulence and outer scale}
\label{sec:parameter-estimation}

We model the fluctuations, $y(t)$, as a zero-mean correlated multivariate Gaussian distribution. If the fluctuations are stationary, the covariance depends only on the delay $\tau = |t - t'|$ and is given by Equation~\ref{eqn:time-covariance} as
%
\begin{equation}
    C(\tau, \sigma^2, \tau_0) = \sigma^2 M_{5/6} (\tau / \tau_0),
    \label{eq:fitting_cov}
\end{equation}
where $\sigma^2$ and $\tau_0$ are parameters to be fitted corresponding to the total variance and the temporal outer scale. For $n$ samples of $y$, the marginal log likelihood $\log \mathcal{L}$ is given by 

\begin{equation}
\label{eqn:log_marginal_likelihood}
    \log \mathcal{L} = - y^\dagger K_{t, t'}^{-1} y - \log \det K_{t, t'},
\end{equation}
with kernel $K_{t, t'} = C \big (|t - t'|, \sigma^2, \tau_0 \big )$. In the interest of comparing points of many different timescales $\tau$, we use 16 azimuths sampled evenly across the approximately 40$^\circ$ scan,
compute the likelihood given the temporal fluctuations for each one, and then taking the total log likelihood to be the sum over all azimuths.\footnote{The projected wind-speed differs somewhat over a $40^\circ$
span in azimuth, which we ignore. This is in line with other approximations such as the frozen 2D sheet approximation and finite azimuth bins. These effects, however, are sub-dominant to the variations in $\tau_0$ and so we simply average over them.} 

\section{Results}
\label{sec:results}

\begin{figure}[tp!]
\includegraphics[width=0.475\textwidth]{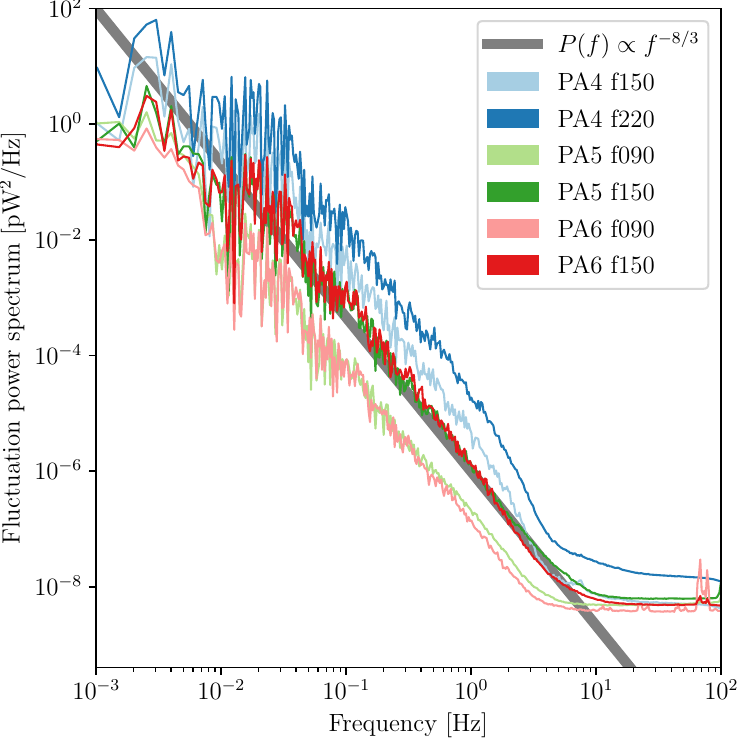}
\caption{The mean power spectral density of all detectors for each band for the TOD shown in Figure~\ref{fig:sample_tod}. Note 
the indication of a flattening spectrum below the outer scale of $\sim 0.02$\,Hz and the -8/3 turbulent spectrum above it. The steepening of the spectrum between 1\,Hz and the ankle at $\sim 5$\,Hz can be explained by the effects of the beam smearing smaller features as described in Appendix~\ref{appendix:two-dimensional}. The form of this spectrum mimics that of the spatial spectrum in Figure~\ref{fig:covariance-triptych}, but scaled by the wind velocity.}
\label{fig:sample_tod_spectrum}
\end{figure}

\subsection{Power spectrum, structure functions and outer scale estimates}

\begin{figure}[htb]
\includegraphics[width=.475\textwidth]{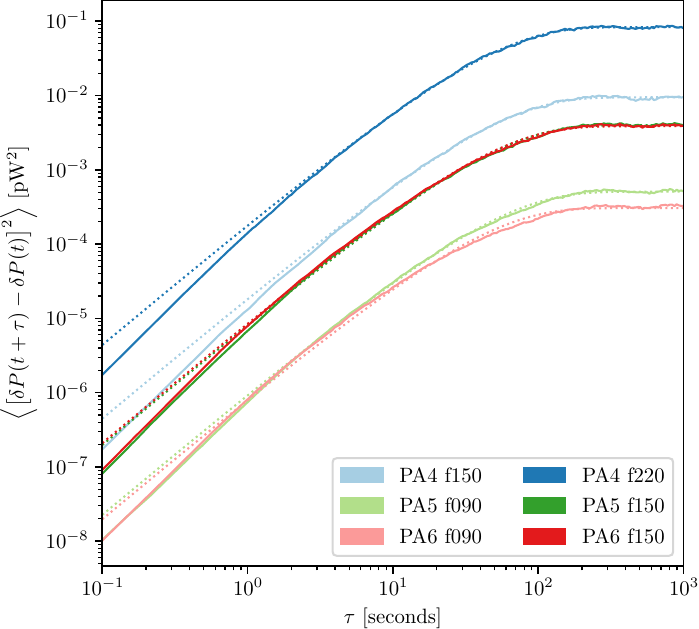}
\caption{The average structure function for each of ACT's bands, contingent on the TOD being selected as atmosphere-dominated. The level of the flat region on the right hand side corresponds 
to the average $2\sigma^2$ from fitting to Equation~\ref{eq:fitting_cov} while the transition from the positive slope to the flat top is determined by $\tau_0$. The dotted lines are predicted structure functions for negligible beams, fitted to the data for separations greater than 10 seconds.}
\label{fig:act-structure-functions}
\end{figure}

Figure~\ref{fig:sample_tod_spectrum} shows the power spectra of the data in
Figure~\ref{fig:sample_tod}. They agree with expectations after accounting for the effects of beam smearing.  Evidence for the outer scale shows up at $f<0.01\,$Hz.
Figure~\ref{fig:act-structure-functions} shows the structure function for each band, averaged\footnote{Because the scaling of the structure functions of different TODs can vary by several orders of magnitude, an explicit average will be most sensitive to TODs in which atmospheric fluctuations are the strongest. In order to probe the average shape of the structure function, we take a weighted average to normalize each structure function with respect to its variance.} over all TODs that were designated as atmosphere-dominated ($a^2 \geq 10)$. The key feature is the flattening for $\tau \gtrsim 100$s, beyond which there are no additional contributions to the variance. 
For each average structure function, we overlay a best fit of the structure function as predicted by the Matérn covariance and accounting for the beam profile and water vapor scale height (see below) as described in more detail in Appendix~\ref{appendix:two-dimensional}.
Figure~\ref{fig:outer_scales} shows the distribution of $\tau_0$ from the fits for PA6 f150 which is representative of the other arrays. In the frozen transport approximation
the spatial length scale differs from the temporal length scale by a factor of the wind speed. 
Thus, to find the outer scale $L_0$ we multiply by the wind speed derived from ERA5 \citep{ERA5}. 
As shown in \cite{morris/etal:2022}, there is an excellent agreement between the wind speed derived from ACT with that derived from the absolute humidity weighted velocity profile in ERA5.


\begin{figure}[hbt]
\includegraphics[width=.475\textwidth]{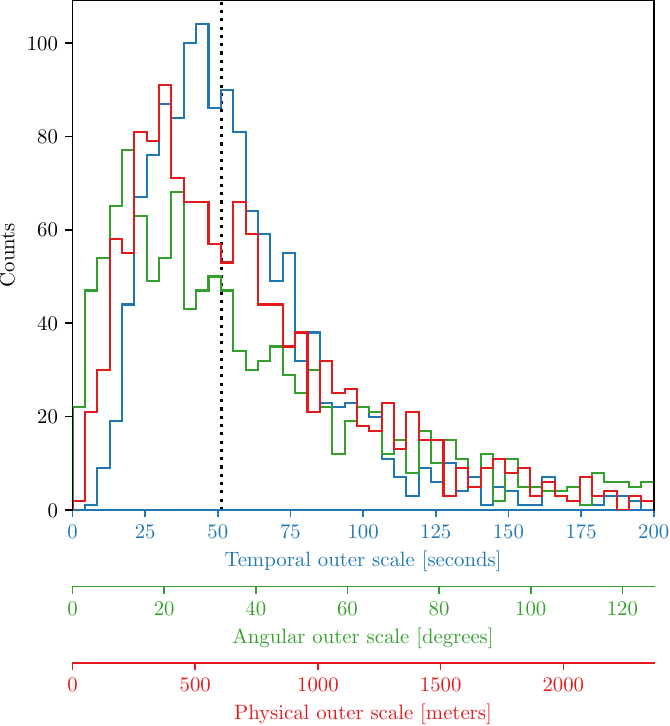}
\caption{The distribution of temporal, angular, and spatial outer scales for ACT. We derive the spatial outer scale by scaling the temporal outer scale $\tau_0$ by the wind speed according to Equation~\ref{eqn:time-covariance} using the wind velocity derived from ERA5. Similarly, We derive the angular outer scale by scaling $\tau_0$ by the angular wind velocity estimated from ACT data. The vertical dashed line coincides with the median of all three distributions.}
\label{fig:outer_scales}
\end{figure}

We also show the distribution of the angular outer scale, computed by scaling $\tau_0$ by the angular velocity of the atmosphere estimated directly from ACT following the method in \cite{morris/etal:2022}. We see in
Figure~\ref{fig:outer_scales} 
that $L_0$ is comparable to the scale height of the water vapor. While not surprising, simultaneously accounting for both the scale height and outer scale leads to a somewhat modified form of the predicted structure function shown in Figure~\ref{fig:act-structure-functions}. We discuss this further in Appendix~\ref{appendix:matern}.

\subsection{Comparison of fluctuations in ACT to APEX}

For all hours when both ACT and the APEX radiometer were observing between 2017 and 2019, we compile the RMS minute-to-minute fluctuations in their respective time 
streams.\footnote{Even though ACT scans back and forth, there are several pairs of samples at the same azimuth but separated in time by 60 seconds. We use the mean squared difference of these pairs to estimate the RMS of minute-to-minute fluctuations.}
We scale ACT data to a common elevation of $45^\circ$, and scale by the proportion of atmospheric variance to the total variance derived using the method in Section~\ref{sec:tod-decomposition}.

\begin{figure*}[tp!]
\includegraphics[width=0.95\textwidth]{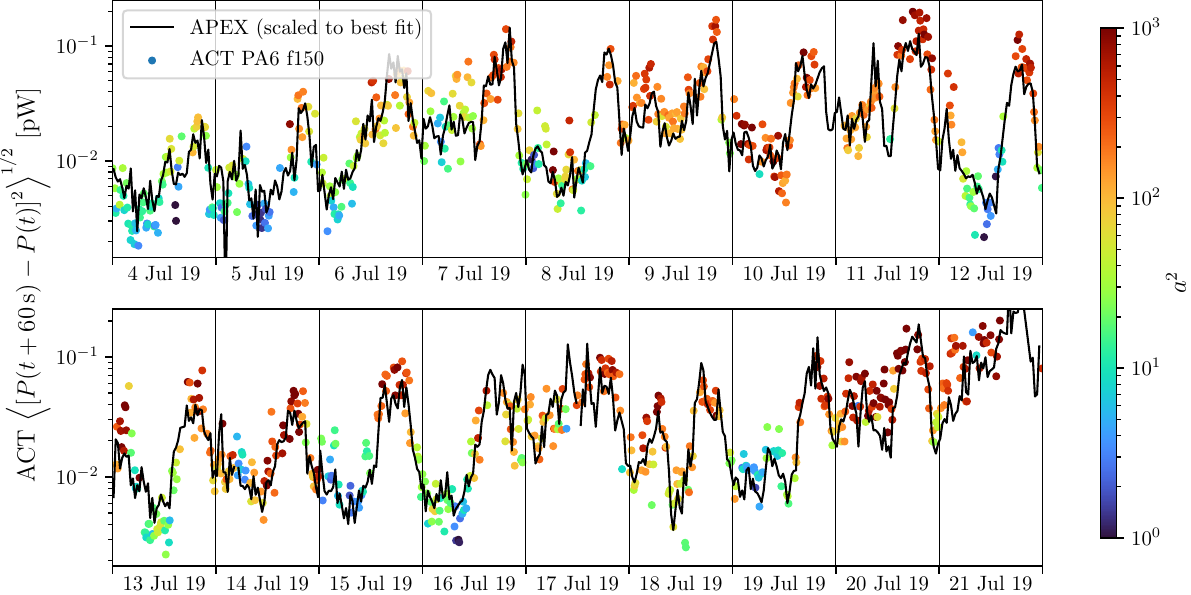}
\caption{Time-ordered variation over several days in the level of minute-to-minute power fluctuations measured by ACT's PA6 150 GHz band. Each point represents an 11-minute TOD and is colored by the fitted ratio of atmospheric variance to systematic variance, which tends to be lower when atmospheric fluctuations are lower. TODs with stronger fluctuations tend to be more atmosphere-dominated. Superimposed are half-hourly water vapor fluctuation levels measured by APEX scaled to best fit, which shows excellent agreement. Both measures exhibit a diurnal variation of more than an order of magnitude.}
\label{fig:act_apex_time_ordered}
\end{figure*}

\begin{figure*}[tp!]
\includegraphics[width=0.95\textwidth]{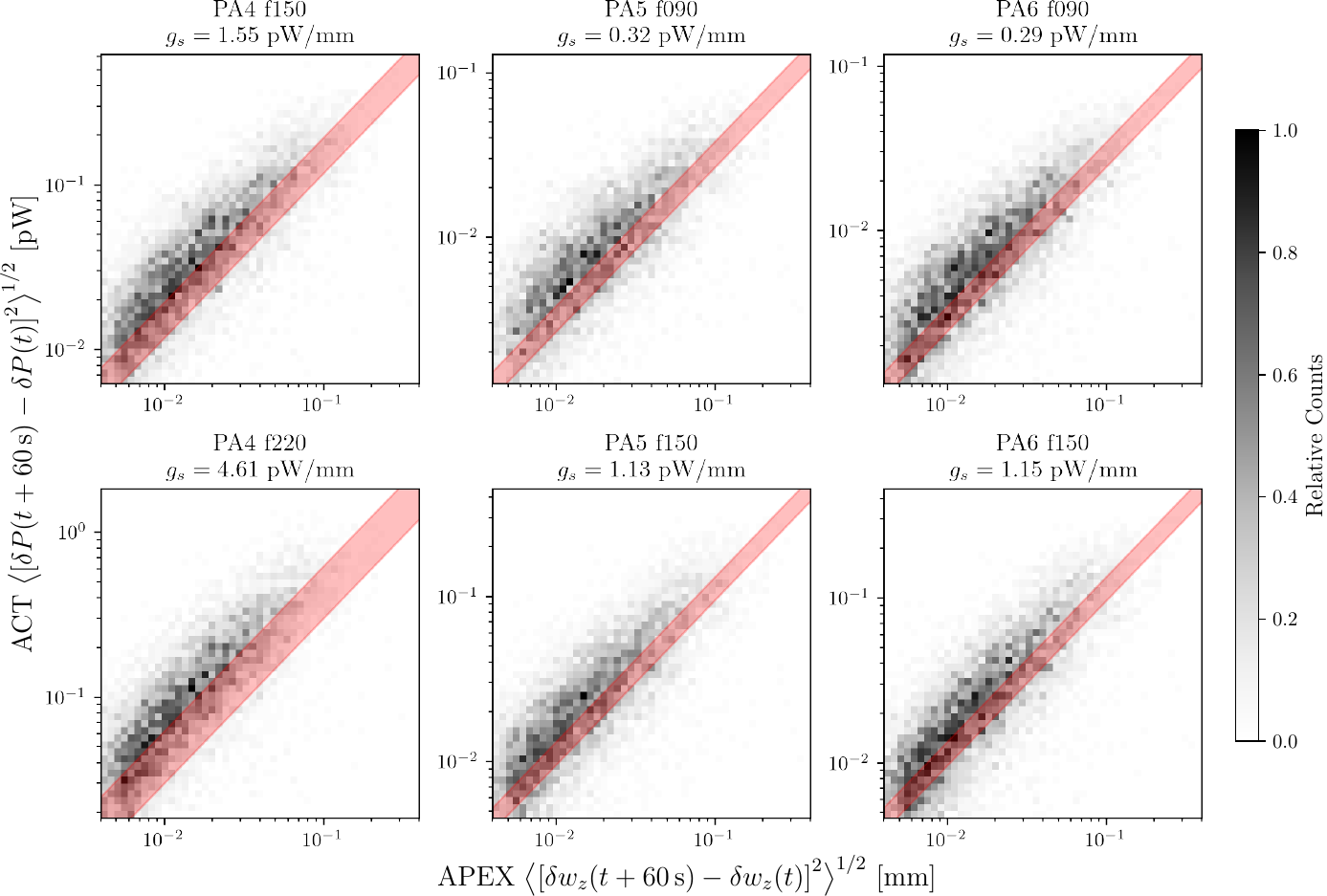}
\caption{The correlation between minute-to-minute fluctuation levels for APEX and for all detectors in each ACT band for all observations in 2019. Again, ACT is scaled to $\theta_{el}=45^\circ$ and the APEX values are for the zenith. 
Varying the lower cutoff for APEX water vapor fluctuations changes the best fit gain; the red band represents the range of best fit gains for cutoffs
between 0.02 mm (top) and 0.1 mm (bottom). These are consistent with predicted responses, as shown in Figure~\ref{fig:fluctuation_spectrum}.}
\label{fig:act_apex_scatter}
\end{figure*}

Figure~\ref{fig:act_apex_time_ordered} shows the level of fluctuations in ACT and APEX varying together over several days. Most of the change in the level of fluctuations is driven by diurnal variations in weather in the Chajnantor region. Figure~\ref{fig:act_apex_scatter} shows the correlation of fluctuation levels in APEX and ACT for atmosphere-dominated TODs ($a^2 \geq 10$), with good agreement between the two. 

For each detector-season, we estimate the response to water vapor fluctuations by comparing ACT to APEX. For each TOD we have, in principle,
\begin{equation}
    \big \langle \delta P^2 \big \rangle^\frac{1}{2} = g \big \langle \delta w_z^2 \big \rangle^\frac{1}{2}.
\end{equation}
We find that both $\langle \delta P^2 \rangle^\frac{1}{2}_t$ and $\langle \delta w_z^2 \rangle^\frac{1}{2}_t$, where the subscript denotes average over a TOD, are distributed roughly log-normally, and hence the distribution of $g$ is log-normal. For a log-normal distribution, the maximum likelihood estimator of $g$ is the median (while the mean is upwardly biased for any nonzero variance due to a heavy tail).
To calibrate ACT's fluctuation to APEX, we account for the difference in height between ACT and APEX ($\sigma_{ACT} = 0.94\,\sigma_{APEX}$, Section~\ref{sec:apex}). This results in a scaled value for $g$ that we call $g_s$. To estimate ACT's efficiency, $\eta$, we divide $g_s$ by the entry for $\partial P/\partial w_z$ in Table~\ref{tab:atmpwr}. Because the resolution of APEX is 0.01~mm, we compute the correlation for RMS fluctuation levels of 0.02~mm and above. As this cutoff level is increased, the computed $\eta$ decreases. The results are reported in Table~\ref{tab:gain_table}.  Based on the variance over seasons, the dispersion of the correlation in Figure~\ref{fig:act_apex_scatter}, the bias associated with the cutoff level, and the uncertainty in the passbands, we estimate the net uncertainty to be roughly 20\%. 

The agreement with previous analyses is reassuring and suggests that calibrating on atmospheric fluctuations is consistent with the more accurate planet calibration. There are, however, a couple of aspects to keep in mind. Although it accounts for most of the variance in the data, our model for separating atmospheric modes is phenomenological; there may be unaccounted for atmospheric variance absorbed into the common mode during the fitting. Additionally, we assume atmospheric fluctuations at APEX and ACT follow a simple scaling even though the sites have different local topography.


There is an overall degeneracy, that we cannot determine, between $\eta$ and any out-of-band passband leakage or an incorrect passband measurement. The quoted values are for the recommended frequency bounds for the integrals over the passbands \citep{hasselfield/etal:2024} and is based on an analysis of the Fourier Transform Spectrometer (FTS) used to measure them \cite{tommy_alford_2024}. 
%
For the 90~GHz bands, there is possible band pass leakage at the 1\% level in the FTS measurements that resembles a second harmonic of the primary passband. Were it real as opposed to a measurement artifact, it would lead to sensitivity to the large atmospheric signal near 180 GHz shown in Figure~\ref{fig:datm_dpwv}. However, there is no clear evidence of sensitivity to this feature.


%
%

Figure~\ref{fig:fluctuation_spectrum} shows the three-year median fluctuations with a 25\% to 75\% range for data with 
$0.8{\rm\,mm}<w_z<1.2\,$mm and $0.015 {\rm\,mm}<\sigma_{w_z}<0.025\,$mm (thicker orange lines). To compare to the model with $\sigma_{w_z}=0.01\,$mm, we divide the measured fluctuations by $2\eta$.
The thinner orange lines show the uncertainty on the median range due to the 20\% uncertainty in $\eta$. Although the efficiency calculation uses the same values for 
$\partial P/\partial w_z$ as the model, the $g_s$s that enter into $\eta$ are derived from a wide range of fluctuations in $w_z$. The green lines show the uncertainty on the median range due to the uncertainty in $\eta_{Choi}$. As seen in the figure and Table~\ref{tab:gain_table}, there is broad consistency between the atmosphere-determined $\eta$ and the planet-derived efficiencies, 
$\eta_{Choi}$, determined by comparing the measured power from Uranus to its calibrated spectrum.

\begin{table}
\centering
\begin{tabular}{ |l||c|c|l|l| }
\hline
Band & $\partial P/\partial w_z$ & $g_s$ & $\eta$ & $\eta_{Choi}$ \\
&  pW/mm & pW/mm & &  \\
\hline
PA4 f150 &   3.36 &  1.65  &  0.49 &  $0.33\pm0.13$ \\
PA4 f220 &  15.6 &  4.93  &  0.32 &  $0.34\pm0.12$\\
PA5 f090 &   0.55  &  0.31 &  0.55 &  $0.37\pm0.08$ \\
PA5 f150 &   2.50  &  1.10  &  0.44 &  $0.35\pm0.07$ \\
PA6 f090 &   0.65  &  0.28  &  0.43 &  $0.39\pm0.09$\\
PA6 f150 &   2.67  &  1.14  &  0.43 &  $0.48\pm0.10$\\
\hline
\end{tabular}
\caption{Measured gains and efficiencies for each ACT band averaged over seasons. For 
$\partial P/\partial w_z$ we use the values at 1~mm. The $\eta_{Choi}$ come from \citet{choi/etal:2018, choi2020atacama} and have been corrected to use $\Delta\nu=\int\tau(\nu) d\nu$, as opposed to the Dicke bandwidth. The error bar represents the dispersion amongst detectors.
}
\label{tab:gain_table}
\end{table}


\subsection{Atmospheric fluctuations at 30 and 40 GHz}
\label{sec:low-frequency}

We do not find any significant correlation between APEX PWV fluctuations and ACT fluctuation levels in the f030 and f040 bands (PA7). Figure~\ref{fig:fluctuation_spectrum} shows that these bands have a larger relative sensitivity to fluctuations in air temperature. We relax our assumption that atmospheric power fluctuations for these bands are entirely proportional to PWV, and write power fluctuations in the atmospheric spectrum as 
\begin{equation}
    \delta T_b(\nu)^2 = \Big [ \frac{\partial T_b(\nu)}{\partial w} \Big ]^2 \delta w^2 +  \Big [ \frac{\partial T_b(\nu)}{\partial T_b} \Big ]^2 \delta T_b^2.
\end{equation}
The atmospheric fluctuations in PA7 are considerably smaller than at higher frequencies, as shown in Figure~\ref{fig:fluctuation_spectrum} and Table~\ref{tab:atmpwr}, and therefore more difficult to separate from systematic effects.
At frequencies below $\sim 1\,$Hz, again lower than those used for making maps, the PA7 data are dominated by scan-synchronous pickup that is not on the sky. In addition, the expected level of fluctuations in the APEX data are near its measurement limit.

Based on a robust wind solution, we can identify a clear atmospheric signal about 3\% of the time. This is usually during bad observing conditions when the higher frequencies are problematic enough that they are cut from mapmaking. In these cases, there is no correlation with the APEX fluctuations and the ratio between signals in the f040 and f030 bands is typically larger than 7, much larger than 
expected for water vapor fluctuations as shown below.

This ratio is compatible with the prediction for temperature fluctuations. It cannot be explained by the temperature of the water vapor because that is the subdominant component of emission at 40 GHz. This can be seen in Table~\ref{tab:atmpwr} by subtracting the power at $w_z=0$~mm from that $w_z=3$~mm for PA7 to get the net water content and computing the ratio between bands: $(1.43-1.21)/(0.28-0.15)= 1.7$, far from what we measure.
Thus it appears that we are seeing temperature fluctuations in the oxygen.

There are two confounding effects that limit what we can say about the dominant source of fluctuations in PA7 in general. One is that the synchronous pickup has a signal ratio $\sim 6$ between the f040 and f030 bands, slightly less, but distinguishable from, the $\ge 7$ we see for the atmosphere, and variations in it can dominate the much smaller atmospheric signal. More analysis is needed to project out the synchronous signal over a broader range of atmospheric conditions.  The other is that we cannot rule out some level of out-of-band contamination by the water lines, but in this case we would expect a correlation with APEX which we don't observe. The ratio of power between the f030 and f040 bands is consistent with expectations from scattering by ice crystals \citep{takakura/etal:2019, li2023class, coerver2024spt}, but the ratio to higher frequencies is inconsistent with this explanation.

The scale height for temperature variations is generally 25~km. As the fluctuations are proportional to the gradient \citep{obukhov:1949}, it must be that some local effect, perhaps an inversion layer \citep[e.g.,][]{cortes/etal:2016, bustos/etal:2014}, is producing a steeper gradient to make sufficiently large fluctuations. Unfortunately, the ERA5 data are not sampled finely enough to clearly identify a local inversion layer at, say, 750\,m.

\section{Discussion and Conclusions}
\label{sec:discussion}


This paper focuses on a statistical description of the atmosphere as observed by ACT. In general we find that the fluctuations are well described by Kolmogorov turbulence, in agreement with previous studies, and further that the full set of statistical descriptors 
--the wind speed, variance, power spectrum, structure function, outer scale-- can be extracted directly from the data after accounting for the effects of the beam.

CMB experiments, like ACT, often cut data based on the PWV level. The comparison with APEX shows that it is likely better to cut on a combination of PWV and its variance as there are times with relatively low PWV and high variance, and vice versa.

Atmospheric fluctuations are often used as a continuous beam-filling source for flat-fielding an array of detectors. For large arrays some care is needed as the angular size of the outer scale of turbulence can be smaller than the telescope field of view. Our analysis focused on ACT's data at 90 GHz and greater because of the clear association with water vapor. At 30 and 40 GHz, the correlation with APEX's water vapor radiometer was much weaker suggesting that flat-fielding with the atmosphere in these bands will be more challenging than at higher frequencies.

Additional water vapor radiometers have operated at the ACT and CLASS \citep[e.g.,][]{appel/etal:2019} sites and there are plans for yet more instrumentation \citep{darcy_private:2024} to complement APEX. Given the correlation with APEX reported here along with the agreement between ACT and ERA, characterizing the atmospheric effects across an array of different nearby CMB instruments simultaneously appears achievable.  

\section{Acknowledgments}

This publication makes extensive use of data acquired by the water vapor radiometer (WVR) on the Atacama Pathfinder Experiment (APEX). APEX is operated by ESO through the Max Planck Institute for Radio Astronomy (MPIfR). We are especially appreciative of detailed discussions on the WVR with Carlos De Breuck (ESO APEX Project Scientist). We also acknowledge helpful conversations with Akito Kusaka.

Support for ACT was through the U.S.~National Science Foundation through awards AST-0408698, AST-0965625, and AST-1440226 for the ACT project, as well as awards PHY-0355328, PHY-0855887 and PHY-1214379. Funding was also provided by Princeton University, the University of Pennsylvania, the Wilkinson Research fund, and a Canada Foundation for Innovation (CFI) award to UBC. ACT operated in the Parque Astron\'omico Atacama in northern Chile under the auspices of the Agencia Nacional de Investigaci\'on y Desarrollo (ANID). The development of multichroic detectors and lenses was supported by NASA grants NNX13AE56G and NNX14AB58G. Detector research at NIST was supported by the NIST Innovations in Measurement Science program. Computing for ACT was performed using the Princeton Research Computing resources at Princeton University, the National Energy Research Scientific Computing Center (NERSC), and the Niagara supercomputer at the SciNet HPC Consortium. SciNet is funded by the CFI under the auspices of Compute Canada, the Government of Ontario, the Ontario Research Fund–Research Excellence, and the University of Toronto. The Flatiron Institute is supported by the Simons Foundation.
CS acknowledges support from the Agencia Nacional de Investigaci\'on y Desarrollo (ANID) through Basal project FB210003.
SN acknowledges support the Simons Foundation (award CCA 918271, PBL). 
RD thanks ANID for grant BASAL CATA FB210003 and FONDEF ID21I10236.
We thank the Republic of Chile for hosting ACT in the northern Atacama, and the local indigenous Licanantay communities whom we follow in observing and learning from the night sky.


\appendix

\setcounter{table}{0}
\renewcommand{\thetable}{A\arabic{table}}

\setcounter{figure}{0}
\renewcommand{\thefigure}{A\arabic{figure}}

\begin{figure*}[htb]
\includegraphics[width=0.95\textwidth]{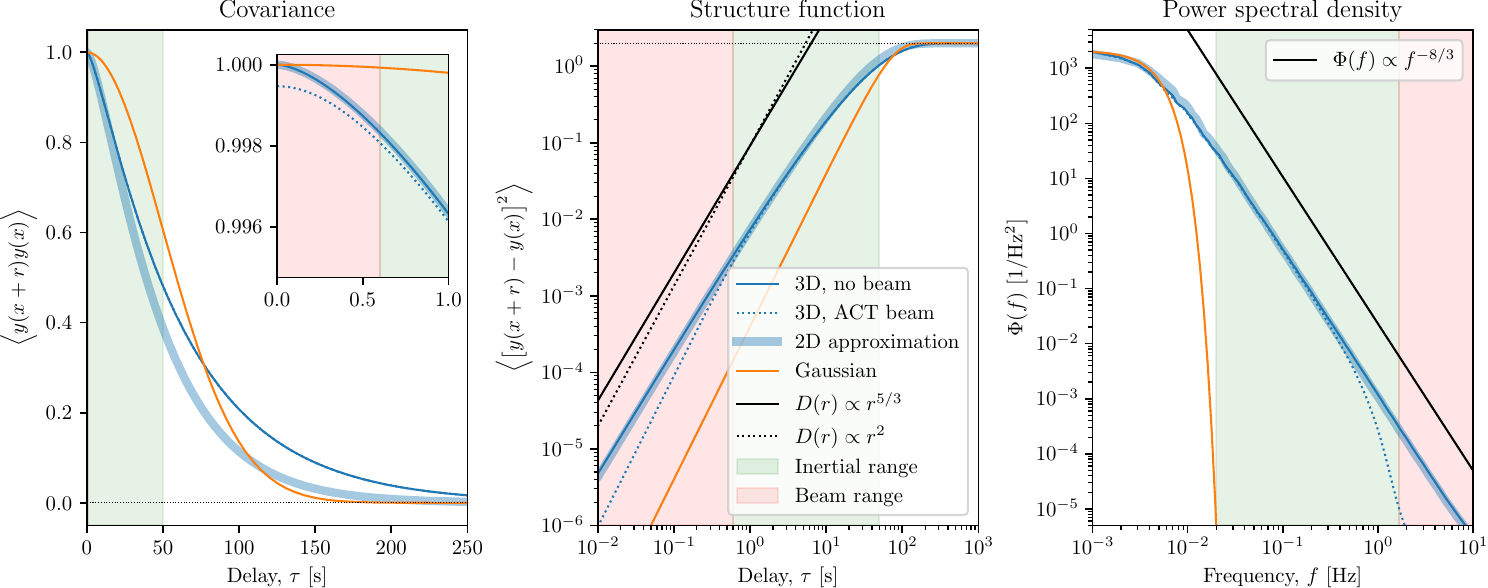}
\caption{The Mat\'ern function, in thin blue, and related statistical measures of atmospheric fluctuations including the result of numerically integrating Equation~\ref{eqn:six-integral} with and without beams, using values of $z_0 = 1\,$km, $L_0 = 500\,$m, and $|v| = 10$. We can see that for ranges outside of the beam size, the covariance admits a two-dimensional approximation with a $5/3$ power law. Accounting for the beam causes a steepening of the structure function from $5/3$ to $6/3$. The power spectrum is obtained as the Fourier transform of the covariance, and shows the same steepening as in the ACT data in Figure~\ref{fig:sample_tod_spectrum}.}
\label{fig:covariance-triptych}
\end{figure*}

\section{Correlations in turbulent atmospheric emission}
\label{appendix:covariance}

\subsection{Derivation of the Matérn covariance function for scale-invariant turbulent distributions}
\label{appendix:matern}

An entirely scale-invariant spectrum is not integrable, and thus physical turbulence cannot be scale-invariant for arbitrarily low wavenumber $\kappa$. We can impose a flat spectrum below some frequency $\kappa_0=1/L_0$ (where $L_0$ is the outer scale of turbulence) with a spectrum of the form
\begin{equation}
\label{eqn:pspec}
    \Phi(\boldsymbol{\kappa}) = \Phi(\kappa) =
    \alpha_0 \big (\kappa_0^{2} + \kappa^2 \big )^{-(\nu+\frac{3}{2})},
\end{equation}
where $\kappa = |\boldsymbol{\kappa}|$ and $\nu=1/3$ for Kolmogorov turbulence. Note that Equation~\ref{eqn:modified-kolmogorov} is the special case of this formula for three dimensions. By the Wiener-Khinchin theorem, the corresponding point-to-point covariance function of the turbulent flow is the three-dimensional Fourier transform of the above
\begin{equation}
\label{eqn:ftpspec}
    C(r) = \mathcal{F}\big[\Phi(\boldsymbol{\kappa})\big](\mathbf{r}) =(2\pi)^3 \int_{\mathbb{R}^3} \Phi(|\boldsymbol{\kappa}|) e^{-i2\pi\boldsymbol{\kappa}\cdot\mathbf{r}}d^3\boldsymbol{\kappa}. 
\end{equation}
Because $\Phi(\kappa)$ is an isotropic and thus has radial symmetry, the three-dimensional Fourier transform is also a radially symmetric function and may be computed as 
\citep[e.g.,][]{grafakos/teschl:2011} %


\begin{multline}
    \mathcal{F}\big[\Phi(\boldsymbol{\kappa})\big](\boldsymbol{r}) = (2\pi)^\frac{3}{2} \int_0^\infty \Phi( \kappa)\big( \kappa / r \big)^\frac{1}{2} J_\frac{1}{2}(\kappa r) \kappa d\kappa \\ =
    2^{-(\nu+\frac{1}{2})} (2\pi)^\frac{3}{2} \alpha_0 \big( r / \kappa_0 \big)^{\nu} \frac{K_{\nu}(\kappa_0r)}
    {\Gamma(\nu+\frac{3}{2})},
\end{multline}
where $J_\frac{1}{2}$ is a Bessel function of the first kind and $K_\nu$ is a modified Bessel function of the second kind.\footnote{See Equation 11.4.44 in \citet{abramowitz/stegun:1970}.}

A convenient choice of the normalization in equation~\ref{eqn:pspec} is for 
$C(r=0)=\sigma^2$ for all $\nu$. 
Because $z^\nu K_\nu(z) \rightarrow 2^{\nu - 1} \Gamma(\nu)$ as $z \rightarrow 0$, this implies
\begin{equation}
    \alpha_0 = \sigma^2 \pi^{-\frac{3}{2}} \kappa_0^{2\nu} 
    \frac{\Gamma(\nu+\frac{3}{2})}{\Gamma(\nu)}.
\end{equation}
Thus we have
\begin{equation}
\label{eqn:two-above}
    C(r) = \sigma^2\frac{2^{1-\nu}}{\Gamma(\nu)}(r\kappa_0)^\nu K_\nu(r\kappa_0) \quad
\end{equation}
and
\begin{equation}
\label{eqn:two-above}
    \Phi(\kappa) = \frac{\Gamma(\nu+\frac{3}{2})}{\pi^\frac{3}{2}\Gamma(\nu )}
    \frac{\sigma^2 \kappa_0^{2\nu}}{(\kappa_0^2+\kappa^2)^{\nu+\frac{3}{2}}}.
\end{equation}
Note that the power spectrum has a different form along one-dimensional translations such as a time delay:
\begin{equation}
    C(\tau) = \frac{2^{1-\nu}}{\Gamma(\nu)}\sigma^2(\tau\omega_0)^\nu K_\nu(\tau\omega_0) \quad
\end{equation}
and
\begin{equation}
    S(\omega) = \sigma^2 \frac{\Gamma(\nu+\frac{1}{2})}{\pi^\frac{1}{2}\Gamma(\nu )}
    \frac{\tau_0}{(1+\omega^2\tau_0^2)^{\nu+\frac{1}{2}}}.
\end{equation}
The normalized Mat\'ern covariance function, 
\begin{equation}
\label{eqn:matern}
    M_\nu(x) = \frac{2^{1-\nu}}{\Gamma(\nu)} x^\nu K_\nu(x),
\end{equation}
where $x=\sqrt{2\nu}r/\rho$, gives a compact generalization of Equation~\ref{eqn:two-above},
$C(r) = \sigma^2M_{\nu}(x)$. Multiplying the length scale $\rho$ by $1/\sqrt{2\nu}$ is a convention that allows $M_{1/2}(x) = \exp(-r/\rho)$
and $\lim_{\nu\to\infty} M_\nu(x) = \exp(-r^2/2\rho^2)$.\footnote{See \url{https://en.wikipedia.org/wiki/Matern\_covariance\_function}.} Note that in this work, $\rho=\sqrt{5/3} /\kappa_0=\sqrt{5/3}L_0$ so that $\kappa_0=1/L_0$ for $\nu=5/6$. For $x \ll 1$, $M_\nu(x)\rightarrow 1-[\Gamma(1-\nu)/\Gamma(1+\nu)](x/2)^{2\nu}$.


\setcounter{table}{0}
\renewcommand{\thetable}{B\arabic{table}}

\setcounter{figure}{0}
\renewcommand{\thefigure}{B\arabic{figure}}

\begin{figure*}
  \centering
\includegraphics[width=1.0\textwidth]{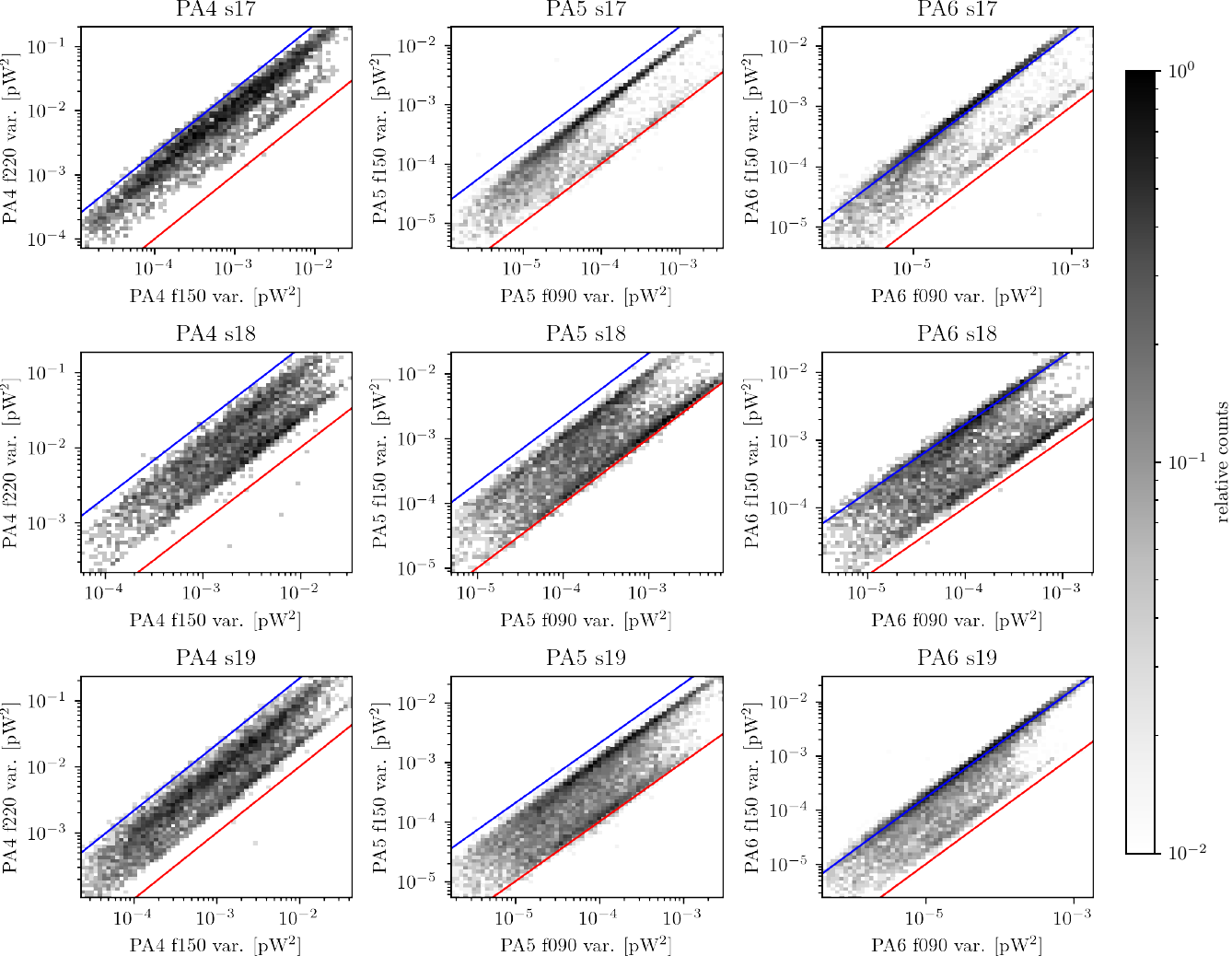}
\caption{The ratio between each pair of bands over all seasons, where each count represents one 11-minute TOD. The ratio is almost entirely dependent on the proportion of common and non-common signals in the time-ordered data. The red line represents a ratio of unity, while the blue line represents the predicted ratio in sensitivities to water vapor fluctuations but not accounting for the relative optical efficiencies between detector arrays.}
  \label{fig:band-scatter}
\end{figure*}

\subsection{Two-dimensional turbulent distributions to the observer.}
\label{appendix:two-dimensional}

To an observer on the ground, the emission of the atmosphere is computed from the integration along the line of sight of a three-dimensional distribution; this can be approximated as a two-dimensional angular distribution. In this section, we relate the general formulation in Section~\ref{appendix:matern} to that observed by ACT. Under the assumption that fluctuations in emission from some region are proportional to fluctuations in water vapor, 
%
the net emission in the optically thin approximation is proportional to the sum of all water vapor fluctuations within the beam: 

\begin{equation}
    \delta w = \int_{\mathbb{R}^3} \delta \rho(\mathbf{x}) B(\mathbf{x}) d\mathbf{x}/\rho_{H_2O},
\end{equation}
where $\delta \rho(\mathbf{x})$ is the fluctuation in water density, $\rho_{H_2O}$ is the density of water, 
$\mathbf{x} = (x, y, z)$ with $d\mathbf{x}$ the volume element, and $B(\mathbf{x})$ is the beam normalized so that
$\int B(\mathbf{x}) dxdy=1$. The integral is performed over a stack of cylinders whose dimensions depend on $z$, the coordinate along the optical axis. We can write the autocorrelation $C(\tau)$ of the water vapor fluctuations as 
\begin{multline}
\label{eqn:six-integral}
    C(\tau) = \big \langle \delta w(t) \delta w(t + \tau) \big \rangle \\ = \bigg \langle \int_{\mathbb{R}^3} \delta \rho(\mathbf{x}, t) B(\mathbf{x}) d\mathbf{x} \int_{\mathbb{R}^3} \delta \rho(\mathbf{x}, t + \tau) B(\mathbf{x}) d\mathbf{x} \bigg \rangle,
\end{multline}
or alternatively as the double integral of each element
\begin{equation}
    C(\tau) = \int_{\mathbb{R}^3} \int_{\mathbb{R}^3} \big \langle \delta \rho(\mathbf{x}, t) \delta \rho(\mathbf{x}, t + \tau) \big \rangle B(\mathbf{x}_i) B(\mathbf{x}_j) d\mathbf{x}_i d\mathbf{x}_j.
\end{equation}
If all elements in the atmosphere are translated at a constant wind velocity $\mathbf{v}$, then we have $\delta \rho( \mathbf{x}, t) = \delta \rho(\mathbf{x} + \mathbf{v} t)$. If the wind velocity is entirely horizontal and the RMS of density fluctuations $\sigma(z) = \langle \delta \rho^2 \rangle^{1/2} $ depends only on height, we can then write this in terms of the spatial covariance as
\begin{equation}
    \big \langle \delta\rho(\mathbf{x}, t) \delta\rho(\mathbf{x}, t + \tau) \big \rangle = \big \langle \delta\rho(\mathbf{x}) \delta\rho(\mathbf{x} + \mathbf{v} \tau) \big \rangle = \sigma^2(z) M_{1/3} \big ( \sqrt{2/5} v \tau \big ), 
\end{equation}
where $v = |\mathbf{v}|$. The quantity $\langle \delta \rho(\mathbf{x}, t) \delta \rho(\mathbf{x}, t + \tau) \rangle$ then becomes equivalent to our spatial covariance from Equation~\ref{eqn:matern}. Consider the case of an infinitely thin beam observing the zenith, such that $B(x, y, z) = \delta(x) \delta(y)$. Modeling $\sigma(z)$ as exponential decay with some scale height $z_0$, our covariance is then

\begin{equation}
    \label{eq:two-integral-covariance}
    C(\tau) = \int_0^\infty \int_0^\infty e^{-(z_i + z_j)/z_0} M_{1/3} \big ( r_\mathrm{eff} \big )  dz_i dz_j,
\end{equation}
where 
\begin{equation}
    r_\mathrm{eff} = \big [ (v\tau)^2 + (z_j - z_i)^2 \big ]^{1/2} / L_0
\end{equation}
The Appendix of \cite{morris/etal:2022} shows that in the limit of small separations $v \tau \ll L_0, z_0$, we recover a two-dimensional model. We make no such assumption in this work; instead, we evaluate Equation~\ref{eqn:six-integral} numerically using quasi-Monte Carlo integration to consider the case of separations comparable to $L_0$ and $z_0$. The results of the evaluation are shown in Figure~\ref{fig:covariance-triptych}. In the regime where separations are much larger than the beam, this leads to a covariance function resembling Equation~\ref{eqn:time-covariance}
\begin{equation}
    C(\tau) \approx \sigma^2M_{5/6}(uL_0^{-1}\tau),
\end{equation}
another Matérn covariance albeit with shape parameter $\nu = 5/6$. The validity of this as an approximation to Equation~\ref{eq:two-integral-covariance} depends on the relation between the outer scale of turbulence and the scale height of the atmosphere. The approximation is generally valid when the $z_0 > L_0$, albeit with an adjustment in the corresponding two-dimensional outer scale. This adjustment vanishes as $z_0 \rightarrow \infty$. Using a scale height of $z_0 = 1.5\,$km and an outer scale of 500\,m, this corresponds to a $\sim 10\%$ adjustment in the 2D vs. 3D outer scale. Throughout this paper, however, we report the outer scale as it parameterizes the two-dimensional model.

\section{TOD decomposition}
\label{appendix:tod-decomposition}

\begin{figure*}
  \centering
\includegraphics[width=0.75\textwidth]{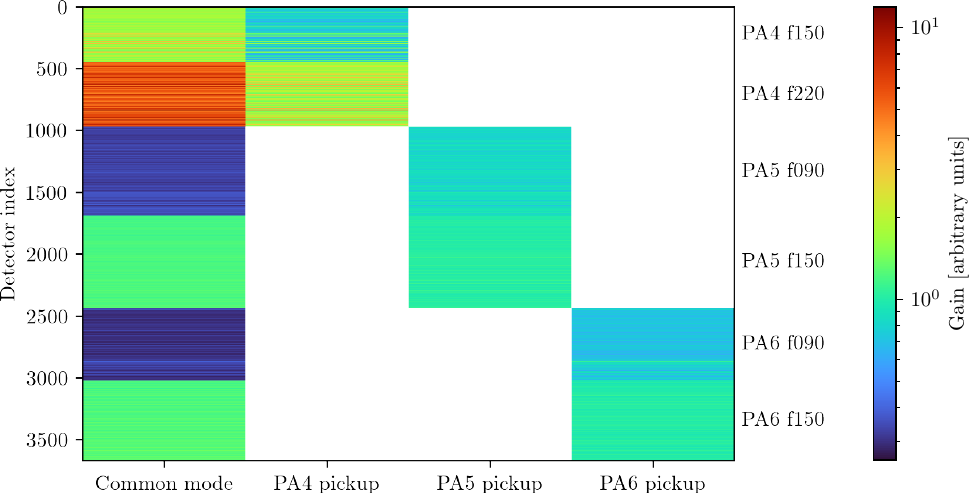}
\caption{The fitted gain matrix for the ACT array for the 2019 observing season (May 2019 - January 2020).}
  \label{fig:gain-matrix}
\end{figure*}

\subsection{Common and non-common modes}

Figure~\ref{fig:band-scatter} shows the ratios in variance in TODs between bands. There are two separate populations visible at the top and bottom of the bands (with intermediate cases in between) corresponding to two different ratios. Atmospheric fluctuations are at the upper end of each band. We posit that each ACT detector timestream is the sum of a mode common to all detectors and a mode common all detectors on the same optics tube, modulo some scaling to each mode. With one common mode and three optics tubes, we can model all detector timestreams for each TOD as a different mixture of 4 distinct timestreams. We represent the time-ordered data as a $(n_\text{detector} \times n_\text{time})$ matrix $\mathbf{P}$. We subtract a cubic spline with a knot spacing of ten minutes for each detector $\mathbf{P}_0$ to be left with the fluctuations $\delta \mathbf{P} = \mathbf{P} - \mathbf{P}_0$. We represent the mixing of common and non-common modes in these fluctuations as a matrix product $\delta \mathbf{P} = \mathbf{G}\,\delta \mathbf{M}$, where $\mathbf{G}$ is an $(n_\text{detector} \times 4)$ matrix of gains, and $\delta \mathbf{M}$ is a $(4 \times n_\text{time})$ matrix representing fluctuations in each mode. We constrain each entry $\mathbf{G}_{i, j}$ such that $\mathbf{G}_{i, j} > 0$ only when $j = 0$ (the common mode) or when $j$ corresponds to the optics tube of detector $i$. The structure of $\mathbf{G}$ is shown in Figure~\ref{fig:gain-matrix}.
If we assume that $\mathbf{G}$ is slowly-varying, we can arrive at a non-degenerate solution by fitting over many TODs simultaneously.

\subsection{Embedding the data}

\begin{figure*}
\begin{center}
\includegraphics[width=1.0\textwidth]{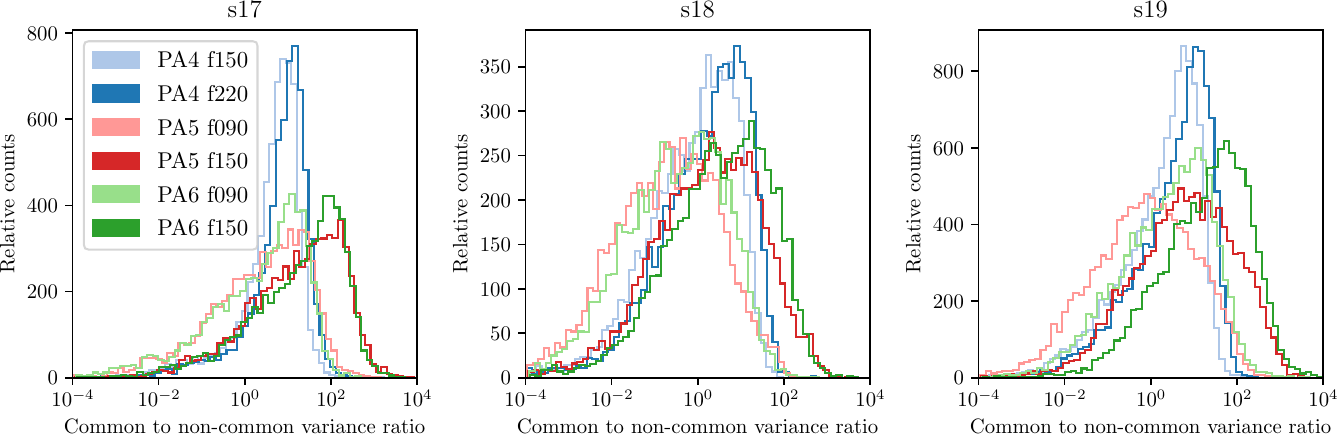}
\caption{A histogram of $a^2$ for each band for each season, showing the ratio between the common and non-common variance of the detectors; a larger $a^2$ indicates a ``purer'' signal free of pickup.}
\label{fig:a-squared_histogram}
\end{center}
\end{figure*}

Fitting over the raw ACT data simultaneously is intractable due to memory limitations. We can, however, condense the data as an ``embedding" into an abstract vector space that preserves only the information necessary to fit Equation~\ref{eqn:tod-decomposition}, i.e., the correlations between the rows of $\delta \mathbf{M}$. Decomposing $\delta \mathbf{P}$ using a singular value decomposition and keeping only the first 4 components gives us
\begin{equation}
    \delta \mathbf{P} = \mathbf{A} \mathbf{B},
\end{equation}
where $\mathbf{A}$ is an $(n_\text{detector} \times 4)$ set of weights and $\mathbf{B}$ is a $(4 \times n_\text{time})$ orthonormal basis. For a given TOD, we have
\begin{equation}
    \mathbf{A} \mathbf{B} = \mathbf{G} \mathbf{M}
\end{equation}
and thus
\begin{equation}
    \mathbf{A} = \mathbf{G} \mathbf{M} \mathbf{B}^+ = \mathbf{G} \mathbf{X}
\end{equation}
where $\mathbf{B}^+$ is the Moore-Penrose pseudoinverse (and thus also the right inverse, because $\mathbf{B}$ is orthogonal). For each TOD, we need only save the weights $\mathbf{A}$, which can easily fit in memory. The interpretation of $\mathbf{X} = \mathbf{M} \mathbf{B}^+$ should be as some embedding in a unique space for each TOD. Because $\mathbf{G}$ varies slowly, the squared norms of each vector in the fitted $\mathbf{X}$ correspond to the relative variance of each fluctuation mode, from which we can compute the relative strength of common and non-common modes, the distribution of which is shown in Figure~\ref{fig:a-squared_histogram} for each band and season.

\subsection{Fitting}

The majority of the time, ACT data is dominated by atmospheric fluctuations common to all detectors. Fitting the above equation in this regime can be overdetermined, which leads to unrealistic solutions for optics tube pickup that are correlated strongly with the atmosphere. This can be solved by applying a prior to $\mathbf{X}$ wherein we down-weigh the probability of common and non-common modes being too correlated. The inner product of two random normalized vectors $\hat{a}, \hat{b}$ in $\mathbb{R}^d$ is distributed as a Beta distribution $\text{B}(x; \alpha, \beta)$ with shape parameters $\alpha = \beta = (d-1)/2$. Our prior is then
\begin{equation}
    p(\mathbf{X}) = \prod_{i={1,2,3}} \text{B} \big ( \hat{\mathbf{X}}_0 \cdot \hat{\mathbf{X}}_i; 3/2, 3/2 \big ),
\end{equation}
where $\hat{\mathbf{X}_i} = \mathbf{X}_i |\mathbf{X}_i|^{-1}$. For each TOD's weights $\mathbf{A}$ and solution $(\mathbf{G}, \mathbf{X})$, we compute
\begin{equation}
    \log \mathcal{L} = \log p(\mathbf{X}) - \log \Big (\frac{\epsilon^2}{2\sigma^2} \Big ),
\end{equation}
where $\epsilon = |\mathbf{A} - \mathbf{G} \mathbf{X}| |\mathbf{A}|^{-1}$ is the normalized residual of the reconstructed embedding. We adopt $\sigma = 0.03$ as it is the median normalized residual after the model is fitted to all ACT data. This model is, in essence, a one-layer neural network that can be implemented in PyTorch and fitted by maximizing $\log \mathcal{L}$ using the backpropagation algorithm with the adaptive momentum (ADAM) optimizer. We fit the model over all TODs in parallel for each season. This produces a solution for $\mathbf{G}$ and $\mathbf{X}$. Despite the degeneracy mentioned in the body of the paper, we can glean from $\mathbf{G}$ the relative gain of each detector and from $\mathbf{X}$ the relative strength of each mode (i.e. the ratio of common fluctuations to non-common fluctuations).





\setcounter{table}{0}
\renewcommand{\thetable}{C\arabic{table}}

\setcounter{figure}{0}
\renewcommand{\thefigure}{C\arabic{figure}}

\section{Numerical Results}
\label{appendix:results}

\begin{figure*}
\centering
\includegraphics[width=1.0\textwidth]{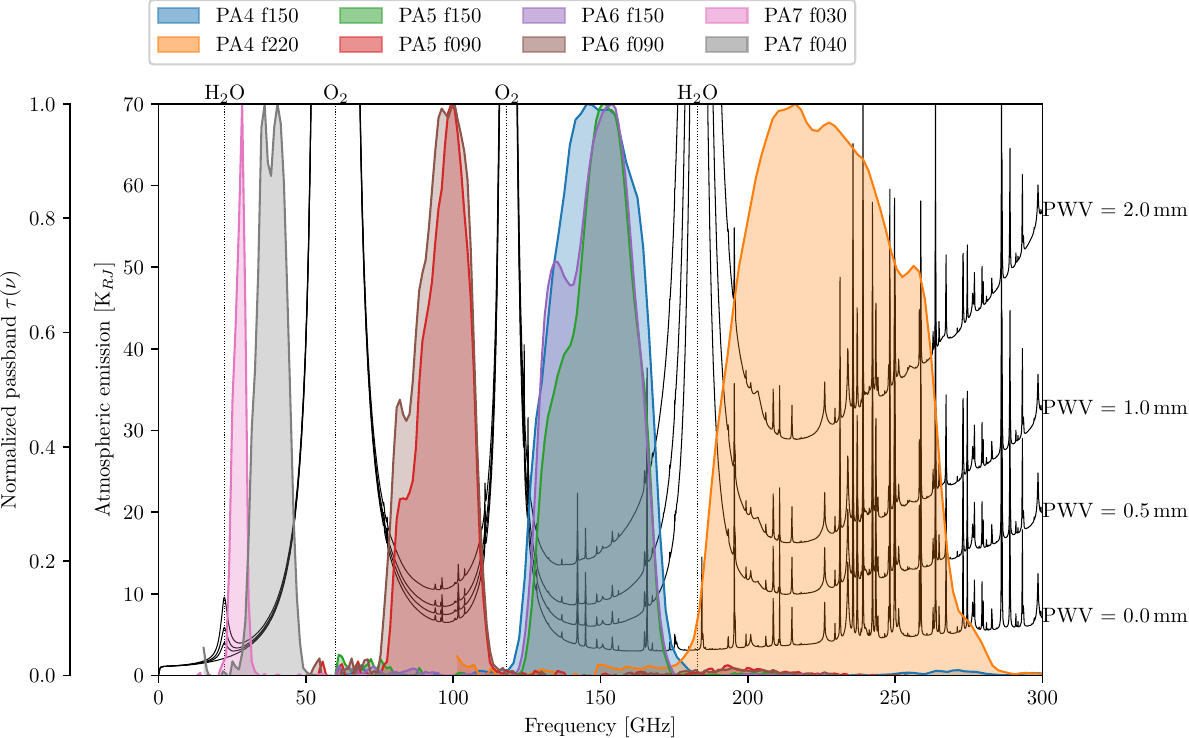}
\caption{The atmosphere's emission spectrum at $45^\circ$ elevation for levels of precipitable water vapor corresponding to $w_z = 0, 0.5, 1, {\rm and~} 2$ mm, from bottom to top respectively. The thin emission lines are from ozone. The normalized ACT passbands for PA4-7, discussed in Section~\ref{sec:act_inst}, are also shown. Spectra for the winter and summer, when adjusted to have the same $w_z$, are the same to the extent we need. These spectra were computed with the \texttt{am} model \citep{am:2019}.}
\label{fig:atmosphere_spectrum_with_passbands}
\end{figure*}

ACT's time streams are calibrated to the CMB. These fluctuations in temperatures may be expressed as fluctuations in picowatts of water vapor by dividing by $\delta T_{CMB}/\delta w_z$. We compute this derivative following the color correction method \citep{planck_IX:2014, hasselfield/etal:2024}:
\begin{equation}
    \frac{\delta T_{CMB}}{\delta w_z} = 
    \frac{\int (I_\mathrm{atm}(\nu )/\delta w_z) \nu^{-2}\tau(\nu )d\nu }{\int b^\prime_\nu \nu^{-2}\tau(\nu )d\nu},
    \label{eq:d20}
\end{equation}
where $I_\mathrm{atm}(\nu )= 2k_B\nu^2T_b(\nu )/c^2$ is the surface brightness of the atmosphere. Here,  
\begin{equation}
b^\prime_\nu =  
\frac{\partial B_\nu(T)}{\partial T}\bigg |_{T_{CMB}}=
\frac{\delta I}{\delta T_{CMB} }=
 b_0\frac{x^4e^x}{(e^x-1)^2},
\label{eq:bprime}
\end{equation}
where $x=h\nu/k_BT_{CMB}$, $B_\nu(T)$ is the Planck function,  and $b_0=2k_B^3T_{CMB}^2/(hc)^2=9.905\times10^{-19}$W/Km$^2$Hz. For the atmosphere it is more natural to work in the change in brightness temperature as reported in  
Table~\ref{tab:atmpwr}. To convert to fluctuations relative to the CMB, multiply the 
derivatives in Table~\ref{tab:atmpwr} by $\delta T_{CMB}/\delta T_{RJ}=(1.02, 1.04, 1.26, 1.27, 1.71, 1.72, 1.73, 3.23)$ for PA7 f030 through PA4 f220 in the same order therein.

The effective frequencies \citep{hasselfield/etal:2024} can be informative but assigning one to the total atmospheric power in a band is problematic because the bands are flanked on both sides by emission lines and so there are often two solutions, one below the band center and another above. However, they are well behaved for the fluctuation spectrum except for the f220 band; we report them in Table~\ref{tab:atmpwr}. \citet{ward/etal:2018} further consider the effects of passband uncertainty on an atmospheric signal in regards to a cosmological analysis. 

\begin{table*}
\begin{center}
\begin{tabular}{ l||lllll }
 \hline
 Channel & $w_z$  & $w_z$ & $w_z$ & $w_z$ & $w_z$ \\
 \hline
   & 0 mm &0.5\,mm & 1.0\,mm & 2.0\,mm & 3.0\,mm\\
 \hline
PA7 f030  ($\nu_c =27.4$\,GHz, $\Delta\nu=4.3$\,GHz)    &   & \\
P (pW) & 0.15 &  0.17  & 0.19 &  0.23   &0.28\\
$\partial P/\partial w_z$ (pW/mm) & &   0.042 &  0.042 ($\dagger$) &  0.042 &  0.043\\
$\partial {\rm T_{RJ}}/\partial w_z$  (K/mm) ($\nu_e =20.1$\,GHz ) & &0.70 &  0.71 &  0.72 &  0.73\\
\hline
PA7 f040   ($\nu_c =38.4$\,GHz, $\Delta\nu=12.3$\,GHz) && & \\
P (pW) & 1.21 & 1.25   &1.28 &  1.35  & 1.43\\
$\partial P/\partial w_z$ (pW/mm)&    & 0.068 &  0.069 (0.012) &  0.073 &  0.077\\
$\partial {\rm T_{RJ}}/\partial w_z$  (K/mm) ($\nu_e =39.8$\,GHz) && 0.40 &  0.41 &  0.43 &  0.46\\
\hline
PA6 f090 ($\nu_c =95.3$\,GHz, $\Delta\nu=23.1$\,GHz) &&  & \\
P (pW) & 2.44 &2.77   &3.09 &  3.75  & 4.45\\
$\partial P/\partial w_z$ (pW/mm)  & &  0.63 &  0.65 (0.020)  &  0.68 &  0.72\\
$\partial {\rm T_{RJ}}/\partial w_z$  (K/mm) ($\nu_e =96.9$\,GHz) & &1.97 &  2.03 &  2.14 & 2.25\\
\hline
PA5 f090 ($\nu_c =96.5$\,GHz, $\Delta\nu=19.1$\,GHz)&  & & \\
P (pW) & 1.91 &  2.19   &2.46 &  3.02  & 3.61\\
$\partial P/\partial w_z$ (pW/mm)&   &  0.53 &  0.55 (0.020) &  0.58 &  0.61\\
$\partial {\rm T_{RJ}}/\partial w_z$  (K/mm) ($\nu_e =95.6$\,GHz) & &2.03 &  2.09 &  2.20 & 2.31\\
\hline
PA6 f150 ($\nu_c =147.9$\,GHz, $\Delta\nu=31.1 $\,GHz) & &   & \\
P (pW) & 1.67 &3.00   &4.33 &  7.04 &  9.80\\
$\partial P/\partial w_z$ (pW/mm) &  & 2.64 &  2.67 (0.032) &  2.73 &  2.79\\
$\partial {\rm T_{RJ}}/\partial w_z$  (K/mm) ($\nu_e =151.4$\,GHz) & &6.16 &  6.24 &  6.38 & 6.50\\
\hline
PA4 f150 ($\nu_c =148.5$\,GHz, $\Delta\nu=36.7 $\,GHz)  &&  & \\
P (pW) & 2.16 &3.86   &5.54 &  8.91 & 12.3\\
$\partial P/\partial w_z$ (pW/mm) & &  3.35 &  3.36 (0.041) &  3.39 &  3.42\\
$\partial {\rm T_{RJ}}/\partial w_z$  (K/mm) ($\nu_e =154.0$\,GHz) && 6.60 &  6.62 & 6.69 & 6.76\\
\hline
PA5 f150 ($\nu_c =149.3$\,GHz, $\Delta\nu=28.1 $\,GHz) && &\\
P (pW) &1.39 & 2.64   &3.89 &  6.41  & 8.98\\
$\partial P/\partial w_z$ (pW/mm) & &   2.48 &  2.50 (0.033) &  2.54 &  2.59\\
$\partial {\rm T_{RJ}}/\partial w_z$  (K/mm) ($\nu_e =153.3$\,GHz) & &6.40 &  6.45 &  6.56 & 6.67\\
\hline
PA4 f220 ($\nu_c =226.7$\,GHz, $\Delta\nu=66.6 $\,GHz) &&&\\
P (pW) &5.59 &  14.6 & 22.6 & 37.6 & 51.6\\ 
$\partial P/\partial w_z$ (pW/mm) & &  16.8 & 15.6 (-0.021) & 14.4 & 13.7\\
$\partial {\rm T_{RJ}}/\partial w_z$ (K/mm) ($\nu_e =226.5$\,GHz) & &18.1 &  16.8 &  15.5 & 14.7\\
\hline
\end{tabular}
\caption{\label{tab:atmpwr} The top line for each band shows the CMB effective frequency and nominal band width for comparison with the effective frequency of atmospheric emission at 1 mm PWV in the bottom row. The f030 band $\nu_e$ is pulled low by the 19 GHz water line. A typical uncertainty for $\nu_e$ is 2 GHz. The numbers in parentheses are the fractional change per GHz. For example, if the PA4 f150 band is shifted up by 1 GHz, $\partial P/\partial w_z$  changes by $3.36\times(0.041)=0.14$.  
$^\dagger$At 30 GHz, where the frequency dependence is strongest,
for $\delta\nu_e=(-2,-1,0,1,2)$~GHz, $\partial P/\partial w_z$ $=(0.078,0.057,0.042,0.034,0.029)$~pW/mm.
All values are integrated through the air column at $45^\circ$ elevation but referenced to $w_z$, the PWV at the zenith.}
\end{center}
\end{table*}

\bibliography{refs_atm2.bib}

\end{document}